\documentclass[twocolumn,floatfix,showpacs,prd,aps,tightenlines,letterpaper,superscriptaddress]{revtex4-1}
\usepackage{amsmath}
\usepackage{graphicx}
\usepackage{psfrag}
\usepackage{verbatim} 
\usepackage[usenames]{color}

\usepackage{dcolumn}
\usepackage{bm}
\usepackage{longtable}
\usepackage[mathscr]{eucal}
\usepackage{mathrsfs}
\usepackage{xspace}
\usepackage{leftidx}
\def\msun{\rm M_{\odot}}
\def\kms{\rm km \, s^{-1}}

\def\simlt{\mathrel{\rlap{\lower 3pt\hbox{$\sim$}}\raise 2.0pt\hbox{$<$}}}
\def\simgt{\mathrel{\rlap{\lower 3pt\hbox{$\sim$}} \raise 2.0pt\hbox{$>$}}}
\def\lsim{\mathrel{\rlap{\lower 3pt\hbox{$\sim$}}\raise 2.0pt\hbox{$<$}}}
\def\gsim{\mathrel{\rlap{\lower 3pt\hbox{$\sim$}} \raise 2.0pt\hbox{$>$}}}

\def\msunpc3{\msun~{\rm {pc^{-3}}}}
\newcommand{\be}{\begin{equation}}
\newcommand{\ee}{\end{equation}}
\def\kms{{\rm\,km\,s^{-1}}}

\def\prx{Phys.\ Rev.\ X}

\newcommand\qmstateproduct[2]{\left\langle#1|#2\right\rangle}

\newcommand{\bea}{\begin{eqnarray}}
\newcommand{\eea}{\end{eqnarray}}
\newcommand{\beq}{\begin{equation}}
\newcommand{\eeq}{\end{equation}}

\usepackage{hyperref}
\hypersetup{
    pdftitle={},
    colorlinks=false,            
}

\begin{document}

\def\fun#1#2{\lower3.6pt\vbox{\baselineskip0pt\lineskip.9pt
  \ialign{$\mathsurround=0pt#1\hfil##\hfil$\crcr#2\crcr\sim\crcr}}}
\def\lap{\mathrel{\mathpalette\fun <}}
\def\gap{\mathrel{\mathpalette\fun >}}
\def\kms{{\rm km\ s}^{-1}}
\def\vk{V_{\rm recoil}}

\title{Hybrid waveforms for generic precessing binaries for gravitational-wave data analysis} 

\author{Jam~Sadiq}
\affiliation{Department of Mathematics, Sukkur IBA University,
 Sukkur, Sindh 65200, Pakistan}
\affiliation{Center for Computational Relativity and Gravitation,
Rochester Institute of Technology, Rochester, NY 14623, USA}

\author{Yosef~Zlochower}
\affiliation{Center for Computational Relativity and Gravitation, 
Rochester Institute of Technology, Rochester, NY 14623, USA}

\author{Richard~O'Shaughnessy}
\affiliation{Center for Computational Relativity and Gravitation,
Rochester Institute of Technology, Rochester, NY 14623, USA}

\author{Jacob~Lange}
\affiliation{Center for Computational Relativity and Gravitation,
Rochester Institute of Technology, Rochester, NY 14623, USA}

\begin{abstract}
We construct hybrid binary black holes merger waveforms using analytical model waveforms for the early inspiral phase and
numerical relativity waveforms for late inspiral to merger and post merger phases. 
To hybridize analytical and numerical  waveforms, we first perform a 3-dimensional rotation to align the instantaneous orbital planes associated with the two waveforms at some fiducial frequency, we then find appropriate phase and time translations that maximize
the overlap of the two waveforms in a  hybridization interval.
We discuss the accuracy and limitations for  hybrids
constructed by this procedure in the context of LIGO-Virgo-KAGRA observations.
Our goal is to hybridize
waveforms for more generic precessing binaries and construct longer waveforms that are sufficiently accurate for the
parameter estimation techniques for upcoming LIGO observations. 
\end{abstract}

 \maketitle

\section{Introduction}\label{sec:Introduction}

With the first ever detection of gravitational waves of 
merging black hole binaries \cite{Abbott:2016blz}, a new era of gravitational wave astronomy has been opened for new and
upcoming 
gravitational wave detectors, such as advanced LIGO, Virgo, KAGRA and LISA \cite{Dwyer:2015fua, Heitmann:2018how, Akutsu:2018axf, Sesana:2014usa, McNamara:2013bma}. LIGO  and Virgo have already observed gravitational waves from merging compact binaries \cite{LIGOScientific:2018mvr} and will be observing more as the O3  observing run continues. There is an expectation that with current capabilities, gravitational wave detectors will observe tens to hundreds of binary black hole mergers every year \cite{Abbott:2016nhf, LIGOScientific:2018jsj, Gerosa:2019dbe} with  binaries with a total mass of 100 times the mass of the sun being observed at the distances of the order of giga parsecs \cite{Abbott:2016nhf}.

  The detection of gravitational waves requires theoretical waveform templates to match the observed data at the
  gravitational wave detector. This technique is called matched filtering, where a theoretically generated waveform
  signal appropriate for a given source is cross correlated against the observed signals at the detector. Because the
  instrumental noise is a random process, a cross correlation will yield positive signature for any signal that matches
  the template within the detectable band, even if the signal is formally weaker than the noise. 
A similar cross correlation arises when inferring source parameters.   A family of theoretically modeled waveforms that depends on the source parameters, such as the two masses, spins, sky location, orbital eccentricities, etc., allows for parameter estimation techniques to be used to infer the properties of the systems that produced the waves~\cite{Cutler:1992tc}.

  To construct the theoretical templates, one needs to solve the Einstein field equations for generic binary black holes. Analytical weak-field approximation methods, such as post-Newtonian theory, can accurately describe the dynamics of such systems in the early inspiral phase prior to merger. Numerical relativity is crucial for the late inspiral to merger phases.  Both of these techniques have been developed and shown to be very successful in the past decade \cite{Blanchet:2013haa,Centrella:2010mx}. It has been shown that analytical model waveforms have similar accuracies to numerical ones for the early inspiral phase of binary black hole systems but lose their accuracy when the binary separation is small. On the other hand, it is practically prohibitive to use numerical relativity for large binary separations, as the simulation time scales roughly  as $T\sim D^4$, where $D$ is the orbital separation. Because of the computational cost of numerical simulations, most numerical relativity simulations of generic precessing binaries cover relatively few orbits prior to merger. These numerical relativity waveforms can be fused together with analytical model waveforms covering the earlier  stage of inspiral. Such fused waveforms are called \emph{hybrid} waveforms.

Hybrid waveforms have many advantages. They combine the best part of two types of waveforms and can play an important role in the construction of phenomenological waveforms \cite{Sturani:2010yv, Khan:2018fmp} and surrogate waveforms \cite{Varma:2018mmi}.
  
  The hybridization of post-Newtonian waveforms with numerical relativity waveforms has been principally explored  for nonspinning
  binaries, as well as binaries where the spins are aligned or antialigned with the orbital angular momentum.
 These hybrid waveforms were then tested for their accuracies and limitations in Refs.~\cite{MacDonald:2012mp,
   PhysRevD.90.124004, Varma:2016dnf, PhysRevD.93.084019, MacDonald:2011ne, Hannam:2007ik, Ajith:2007xh, Hannam:2010ky}.
 Limited aligned-spin NR hybrids have been  used to interpret LIGO observations \cite{LIGO-O2-Catalog}.
Other studies have also manually constructed hybrids for selected precessing waveforms 
 \cite{2012PhRvD..86j4063S,2016PhRvD..93b4003K,2014PhRvL.113o1101H}.
 While no observations yet reported have strong evidence for precession, as deduced by applying semianalytic templates
 to O1 and O2 observations, 
recent studies have indicated that neglecting precession can significantly impact detections and parameter estimations in upcoming runs~\cite{Harry:2013tca, Chatziioannou:2014bma, Canton:2014uja}. Thus, having precessing waveforms is now crucial. 

  Hybridizing precessing waveforms is a complicated process in comparison to  the hybridization of nonprecessing waveforms.  The reason is that the orbital precession strongly affects the gravitational waveforms by modulating both amplitude and phase. This produces a complex waveform that contains rich information about the binary's parameters. In addition, because the orbital plane precesses one needs to {\it rotate} the analytical and numerical waveforms into some standard frame before hybridizing. In addition, there are also a lack of accurate model waveforms for such configurations and work is in progress. Here, we describe a new code that both automates and extends a procedure first described in~\cite{Schmidt:2012rh} to hybridize precessing waveforms, as well as provide an analysis of the various sources of hybridization error.

This paper is organized as follows. In Sec.~\ref{sec:techniques}, we describe the techniques we use to construct the hybrid waveforms. In Sec.~\ref{sec:results} we construct hybrids for two precessing and two nonprecessing systems. In Sec.~\ref{sec:analysis}, we analyze the accuracy of our hybrids. Finally, in Sec.~\ref{sec:discussion}, we review our results and discuss the advantages and limitations of our procedure.

\section{Techniques}\label{sec:techniques}

\subsection{CoPrecessing frame}
\begin{figure*}
  \includegraphics[width=.8\columnwidth]{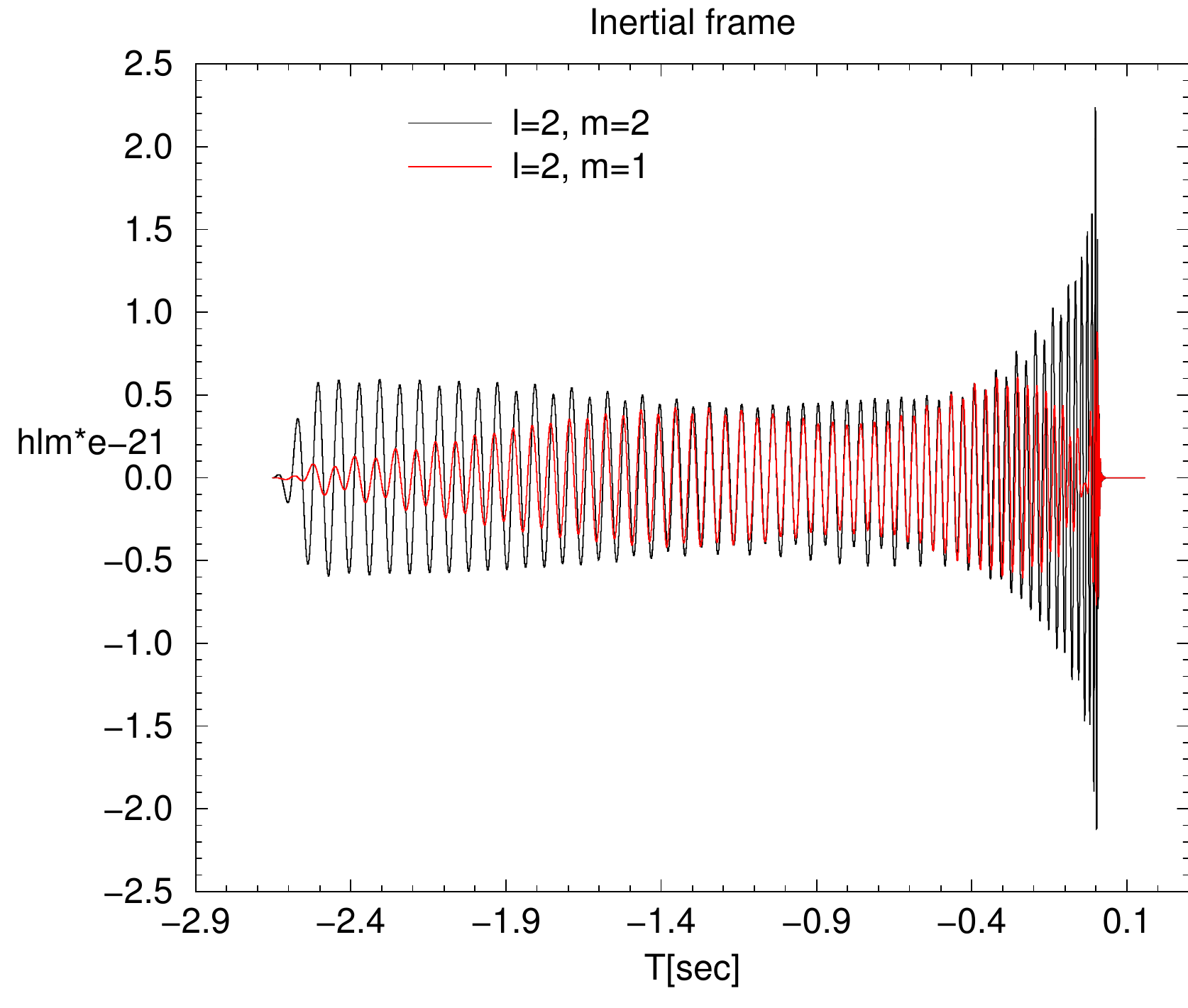}
  \includegraphics[width=.8\columnwidth]{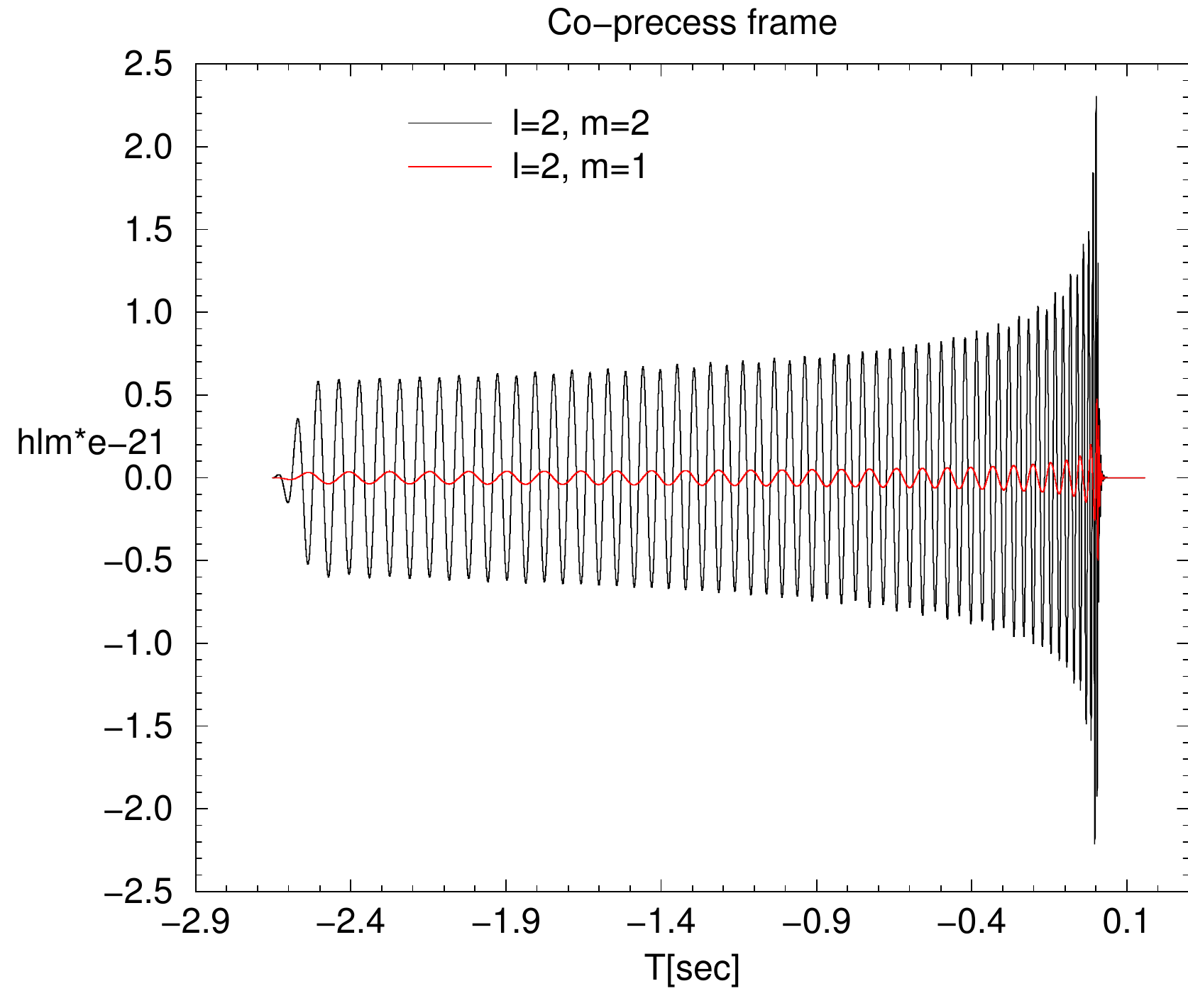}
  \caption{(left) The real part of  $(\ell =2, m=2)$  and  $(\ell =2, m=1)$ modes of a precessing binary black hole (SXS:BBH:0058) with $q=5$, $\chi_1 = (0.5,0,0)$ , $\chi_2 =(0,0,0)$. The  $(\ell =2, m=1)$ contains significant energy and is important for LIGO data analysis for gravitational waves from such precessing binaries\cite{mroue_abdul_2019_3312153}.
 (right) The corresponding coprecessing frame waveform.
 In the coprecessing frame the precessing binaries behaves like a nonprecessing binary with $(\ell =2 , m=2)$ mode being the dominant mode of radiation.}
\label{fig:InertialWaveform2mmodes}
\label{fig:CoprecessingWaveform2mmodes}
\end{figure*}

The dynamics of binary black holes is significantly affected by the
spins of individual components. The details of how gravitational radiation is
produced also depends on the spin of the two compact objects. The spin of a body 
thus imprints itself on the gravitational wave signal. When the spins of either 
one or both compact objects  are not aligned with the orbital plane axis,  both 
the orbital plane itself and the individual spins can precess. This precession 
can impart interesting modulations on the gravitational-wave signal. The $(\ell=2,m=\pm2)$ quadrupolar
mode is not necessarily the most dominant mode as energy is
transferred into other modes, as seen in Fig.~\ref{fig:InertialWaveform2mmodes}.

Due to the effects of precession, the usual procedure for hybridization of nonprecessing waveforms, which amounts to a time and a phase translation of the two waveforms, as has been done in ~\cite{MacDonald:2012mp, PhysRevD.90.124004, Varma:2016dnf, PhysRevD.93.084019, MacDonald:2011ne, Hannam:2007ik, Ajith:2007xh, Hannam:2010ky}, is not sufficient to obtain a reasonable hybrid. We solve this problem using the extra step of performing a full 3-dimensional rotation of the two waveforms such that, at a given time, their principle axes align. 
It has been shown that precessing dynamics can be efficiently estimated via two independent procedures.  In first approach described in ~\cite{PhysRevD.84.024046}, a maximization procedure is used to maximize the magnitude of $(\ell=2, m=\pm2)$  modes by Euler rotations.  These rotations align the orbital angular momentum of binary along the z-direction and thus the $(\ell=2, m=\pm2)$ waveform modes become dominant. These two Euler angles can also be efficiently obtained in another approach described in ~\cite{OShaughnessy:2011pmr} which is based on a preferred direction $\mathrm{\hat{V}}$ aligned with the principal axes of tensor $\left \langle \mathcal{L}_{(ab)} \right \rangle$. This tensor is defined by
\begin{equation} 
\left \langle \mathcal{L}_{(ab)} \right \rangle = \frac{\sum_{\ell m m'}  h_{\ell m'}^*h_{\ell m}    \left \langle \ell m' |\mathcal{L}_{(a}\mathcal{L}_{b)}|\ell m \right \rangle   }{\int d\Omega |h|^2},
\end{equation}
where 
$\mathcal{L}_a$ are the rotation group generators and
\begin{equation}
h = \sum_{\ell=2}^\infty \sum_{m=-\ell}^\ell h_{\ell m} {}^{-2}Y_{\ell m}.
\end{equation}
This components of $\left \langle \mathcal{L}_{(ab)} \right \rangle$ can be expressed as:
\begin{align*}
\left \langle \mathcal{L}_{(ab)} \right \rangle &=& \frac{1}{\sum_{lm}|h_{lm}|^2}
\begin{bmatrix}
I_0 + \text{Re}(I_2) & \text{Im} I_2  &   \text{Re} I_1 \\
 \text{Im} I_2    &   I_0 - \text{Re}(I_2) & \text{Im} I_1 \\
  \text{Re} I_1  & \text{Im} I_1  & I_{zz}
\end{bmatrix}
\end{align*}
where,
\begin{eqnarray}
I_2&\equiv&  \frac{1}{2}\,(h|L_+L_+|h) \nonumber \\
&=& \frac{1}{2}\,\sum_{lm} c_{lm}c_{l,m+1} h_{l,m+2}^*h_{lm} \nonumber \\
I_1 & \equiv &(h|L_+(L_z+1/2)|h)  \nonumber \\
&=& \sum_{lm} c_{lm}(m+1/2) h_{l,m+1}^*h_{lm} \nonumber \\
I_0 &\equiv& \frac{1}{2}\left(h| L^2 - L_z^2 |h\right) \nonumber\\
I_0 &=& \frac{1}{2}\sum_{lm} [l(l+1)-m^2]|h_{lm}|^2   \nonumber\\
I_{zz} &\equiv& (h|L_z L_z |h) = \sum_{lm} m^2 |h_{lm}|^2  \nonumber,
\end{eqnarray}
with $c_{lm} = \sqrt{l(l+1)-m(m+1)}$.

Two of the Euler angles are related to principal axes $\hat{V}$ of  the orientation-averaged tensor by 
  \begin{eqnarray*}
 \alpha  =  \cos^{-1}[\hat{v}_z] 
\end{eqnarray*}
\begin{eqnarray*} 
\beta   =  \text{Arg}[\hat{v}_x + i \hat{v}_y]  -\frac{\pi}{2} 
\end{eqnarray*}
The remaining Euler angle can be computed using \cite{Boyle:2011gg} which account for the gradual buildup of transverse phase due to precession and is given by
 \begin{eqnarray*}
 \gamma  = - \int \dot{\alpha} \cos\beta \quad  dt
\end{eqnarray*}

Rotating the waveform using these Euler angles causes the $(\ell=2$, $m=\pm2)$ modes to become dominant. The resulting coprecessing modes are given by
\begin{equation}
h_{\ell m}^R = \sum_{m'} D_{m m'}^\ell(\alpha,\beta,\gamma) h_{\ell m},
\end{equation}
where the Wigner rotation matrix $ D_{m m'}^\ell(\alpha,\beta,\gamma)$ is given by $D_{m m'}^\ell = d_{m m'}^\ell(\beta) e^{i(m\alpha +m'\gamma)}$ with  $ d_{m m'}^\ell(\beta)$ given by
\begin{align}
\label{eq:Wigner}
d^l_{m'm}(\beta)=&\sqrt{(l+m)!(l-m)!(l+m')!(l-m')!} \nonumber \\
&\times\sum_{k}\frac{(-1)^{k+m'-m}}{k!(l+m-k)!(l-m'-k)!(m'-m+k)!}  \nonumber \\
&\times \left(\sin{\frac{\beta}{2}}\right)^{2k+m'-m}\left(\cos{\frac{\beta}{2}}\right)^{2l-2k-m'+m}.
\end{align}

In this rotating frame, the waveform modes behave very similar to those of a nonprecessing binary system, as can be seen in Fig.~\ref{fig:CoprecessingWaveform2mmodes}. 

In the present work, we use fixed rotations to transform the waveforms into an instantaneously coprecessing frame at the start of the hybridization interval
$H^{\mathrm{rot}}_{lm} (t)  = \sum\limits_{m' =-l}^{l} e^{i m' \gamma + i m \alpha} d_{m m'}^{l} (\beta) h_{lm} (t) $.
Here, $(\alpha, \beta, \gamma) $ are angles at the fixed  time, such that, at that time, the orbital planes associated with the two waveforms are aligned.
It is important to note that the rotation angles are constant in time, thus the waveforms  are still in an inertial frame.

\subsection{Hybridization procedure}\label{sec:hybrid_procedure}

The numerical and analytical waveforms are expressed in different gauges and can use different conventions for the polarization. Thus in addition to performing a 3-dimensional rotation to align the waveforms at a fixed time, we have the additional freedom of adding an arbitrary time translation and phase shift to either waveform, and an additional degree of freedom of multiplying the entire waveform by a fixed phase $\Psi$. The choice of time translation can be chosen by aligning the frequency of two waveforms in a hybrid interval. We align the frequency of two waveforms at a reference frequency in the inertial frame. The reference frequency is chosen to be the frequency of the numerical waveform at the start of hybrid interval. 
We then optimize over time translations, phase shifts, and polarization angle using a \emph{Nelder Mead downhill simplex minimization} algorithm, as implemented in Scipy\cite{Scipy:Py}. In order to find the global minimum we optimize using several different initial guesses for the time shift (close to the one obtained from the coprecessing frame) and several choices for phase shifts in $[-\pi, \pi]$. In all cases we found that the ideal choice of $\Psi$ is either 0 or $\pi$ (this is expected because the two standard choices for the polarization differ by $\pi$). The function we optimize is
 \begin{equation*}
	 \Delta =  \mathrm{min}_{t_0, \phi_0} \int_{t_1}^{t_2} \sum_{l, m} \left|H^{\mathrm{NR}}_{lm} (t) -    H^{\mathrm{MODEL}} _{lm}(t-t_0) e^{i(m\phi_0+2\Psi)} \right|   \mathrm{dt}.
\end{equation*}
Here $H^{\mathrm{NR}}_{lm}(t)$ is the NR waveform and $H^{\mathrm{MODEL}} _{lm}(t-t_0)$ is the model waveform shifted in time, and rotated, such that, at the start of the hybridization interval the principle axes of the NR and MODEL waveforms agree. Note that the rotation of the model waveform depends on the value of $t_0$.
After optimizing for $t_0$, $\phi_0$ and $\Psi$, we taper the time domain waveform using a Planck window~\cite{McKechan:2010kp} and then zeropad to the nearest power of two. The tapering at the start of the waveform is done to avoid Gibbs phenomena at the start of waveform. The tapering at the end is done right after the merger happens to avoid issues with errors in the numerical waveforms during the latter part of the ringdown phase.

After obtaining the appropriate phase and time shifts, we construct the hybrid waveforms via
\begin{equation}
  h^{\mathrm{hyb}} _{l m} =   \tau (t) H^{\mathrm{NR}}_{lm} (t) + \left[1 - \tau(t)\right] H^{\mathrm{MODEL}} _{lm}(t-t'_0) e^{i(m\phi'_0+2\Psi')} ,
\end{equation}
 where  $\tau(t)$ is a function that smoothly goes from 0 to 1 in the  hybrid interval and is given by
\begin{align}
\tau(t) =  \left\{
\begin{array}{cc}
      0 & t <  t1 \\
      \frac{1}{2} (1 + \text{cos}\bigg(\frac{\pi (t-t1)}{(t2-t1)} \bigg) & t1\leq t\leq t2 \\
      1 & t >  t2 \\
\end{array} \right.
\label{eq:TransitionFunction}
\end{align}

We implemented our hybridization procedure using Python. As a test of the timing of our code, we hybridized a numerical waveform 15 orbits prior to merger with a model waveform that was 40 orbits longer. We used a hybrid interval containing 12 cycles (6 orbits). Using these data, the optimization took 40 seconds for each choice of initial time and phase offsets.

\section{Results}
\label{sec:results}
\subsection{Configurations}

\begin{table*}[t]
\centering
\begin{tabular}{ |p{2.2cm}| p{0.75cm}|p{3.02cm}|p{3.02cm}| p{1.1cm}| p{1.1cm}| p{0.8cm}| p{1.4cm}| p{1.4cm}|}
 \hline
\multicolumn{8}{|c|}{} \\
 \hline
	Waveform  & q & $\vec{\chi}_1$  &  $\vec{\chi}_2$  & $\mathrm{N_{cycles}}$ & $\mathrm{f_{ref}}$ & $\mathrm{\phi_{ref}}$ & $\mathrm{fhyb_{ref(40)}}$ & $\mathrm{fhyb_{ref(20)}}$\\
 \hline
	SXS:BBH:0056  & 5      &  (0.0,0.0,0.0 ) & (0.0, 0.0, 0.0)  & 56.4  &  14.608 & 0.0  &   &\\
\hline
	SXS:BBH:0047  & 3      &  (0.0,0.0, 0.5 ) & (0.0, 0.0, 0.5)  & 44.5  & 16.37 & 0.0 &   &\\
 \hline
	SXS:BBH:1392  & 1.53  &  (-0.395,0.229,0.168) & (0.354,-0.125,-0.253) & 281.2 & 4.73801 & 0.0 &  15.18 & 21.95  \\

 \hline
	SXS:BBH:1410  & 4      &  (0.239,-0.318,0.244) & (-0.361,0.039,0.289)  & 154.24 &  8.5045 & 0.0   &  17.673 & 25.36 \\
\hline
RIT:BBH:0137  & 2      &  (0.353,0.0,0.353) & (-0.353,0.0,0.3536)  & 63.77 &  11.6455 & 0.0   &  15.27 & 22.37 \\
\hline

\end{tabular}
	\caption{The waveforms used for analysis. The first column gives the identification string of the waveform as provided in the SXS catalog~\cite{SXS:catalog} and RIT catalog~\cite{RIT:catalog}, $q$ is the mass ratio of the binary, $\vec \chi_1$, and $\vec \chi_2$ are the initial dimensionless spin vectors of the two components, $\mathrm{N_{cycle}}$ is the number of cycles in the $(\ell=2, m=2)$ mode of the waveforms, $\mathrm{f_{ref}}$ is the reference frequency (in Hertz) used to construct the corresponding approximant waveform and $\mathrm{\phi_{ref}}$ is the reference phase which is taken to be zero in all cases. The last two columns shows the reference frequency at the start of hybrid interval for the hybrid constructed using 40 and 20 cycles of the numerical waveforms. These values correspond to $M_{\text{tot}} = 70 M_{\odot}$. Note that the coordinates are chosen such that the two components of the binary lie on the $x$-axis (with the large mass component on the $+x$-axis), and the orbital angular momentum is initially along the $z$direction.}
\label{tab:AllwaveformUsed}
\end{table*}

We constructed hybrids for a  few binary black-hole systems with different properties. We show results for five cases (three precessing, two nonprecessing). In order to hybridize our waveforms consistently, we perform all hybridizations on waveforms corresponding to binaries with a total mass of $M_{\text{tot}} = 70 M_{\odot}$. It is only after hybridizing that we rescale to different masses. In Table~\ref{tab:AllwaveformUsed}, we provide the mass ratio and initial spin configurations for each of the five test configurations. Note that four of the NR waveforms were obtained from the SXS catalog  \cite{SXS:catalog, Mroue:2013xna} and the fifth was obtained from the RIT catalog~\cite{Healy:2019jyf, Healy:2017psd}. For the model waveforms,
we use the post-Newtonian waveforms from the spin-Taylor T4 approximant based on \cite{Ajith:2011ec, Boyle:2014ioa, Marsat:2013caa, Bohe:2015ana, Bohe:2012mr, Marsat:2014xea, Levi:2014gsa}. The waveforms are generated from lalsuite \cite{lalsuite}. For the two nonprecessing cases, we also use waveforms from the EOB models~\cite{Pan:2011gk, Pan:2009wj, Taracchini:2012ig, Taracchini:2013rva, Pan:2013rra, PhysRevD.98.084028}. In this case, we use the SEOBNRv4HM \cite{PhysRevD.98.084028} implementation in lalsuite for the nonspinning and for the spinning case. For brevity, we refer to the spin-Taylor T4 approximant as the {\it PN} waveform and the EOB approximant as the {\it EOB} waveforms.

The first system we hybridized was a nonspinning binary system with mass ratio $q=5$. Here we used the SXS:BBH:0056 waveform from the SXS catalog~\cite{mroue_abdul_2019_3312192} and the corresponding spin-Taylor T4 and SEOBNRv4HM approximants, as obtained from~\cite{lalsuite}. We then hybridized a spinning, but nonprecessing system, with $q = 3$ and $\chi_1 = (0,0,0.5)$ and $\chi_2 = (0,0,0.5)$. Here we used the SXS:BBH:0047 waveform from the SXS catalog \cite{mroue_abdul_2019_3311869} and both the SEOBNRv4HM and  spin-Taylor T4 waveforms (again, as obtained from ~\cite{lalsuite}). Finally, we hybridized two mildly precessing binary black hole systems. These were SXS:BBH:1392~\cite{mroue_abdul_2019_3311869, sxs_collaboration_2019_3315664}, which has  $q =1.513$ and initial spins of $\chi_1 = (-0.3955, 0.229, 0.168)$ and $\chi_2 = (0.35401, -0.125, -0.253)$. The other precessing waveform was SXS:BBH:1410~\cite{mroue_abdul_2019_3311869, sxs_collaboration_2019_3315705}, which has $q =4.0$ and initial spins $\chi_1 = (0.2399,-0.3186, 0.2448)$ and $\chi_2 = (-0.3612, 0.0393, 0.2897)$. 
In both of these precessing cases, we used the  spin-Taylor T4 approximant  with the same initial parameters as the numerical waveforms. In the next section, we show the numerical and analytical model waveforms before our hybridization procedure and after it, and then compute the mismatch as function of total mass.
We analyze the waveforms and discuss different hybrid errors and issues in the analysis section.

\subsection{Hybrid waveforms}

When constructing the hybrids, we need to align the numerical and analytical waveforms. This alignment consists of a time translation and, in general, a full 3-dimensional rotation of one or both waveforms. In the nonprecessing case, a rotation by an angle $\phi$ about the $z$-axis is equivalent to a phase shift of an $m$-mode by $e^{m\phi}$. 

	For the nonspinning configuration (SXS:BBH:0056) we construct the hybrid using the  corresponding post-Newtonian waveforms using the spin-Taylor T4 approximant. We constructed hybrids of all modes except the $m=0$ modes. We compare this hybrid with the available modes of the same system using the SEOBNRv4HM approximant, which has the ($\ell=2 \, , m=\pm2$), ($\ell=2 \, , m=\pm 1$), ($\ell=3\, , m=\pm3$), ($\ell=4 \, ,m=\pm4$) modes. 

\begin{figure*}
\includegraphics[width=.85\columnwidth]{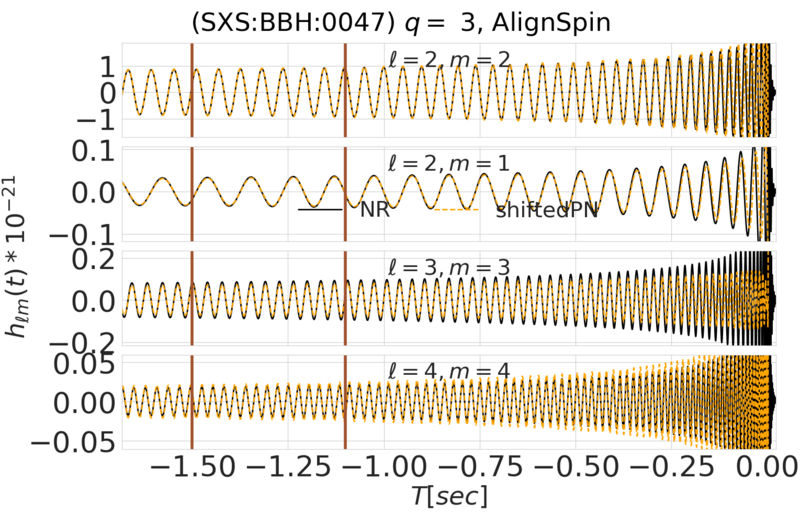}
\includegraphics[width=.85\columnwidth]{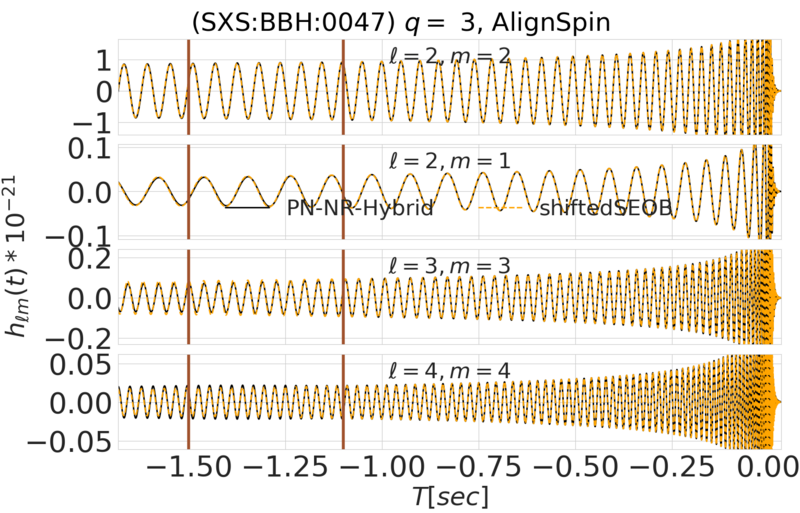}

\includegraphics[width=.85\columnwidth]{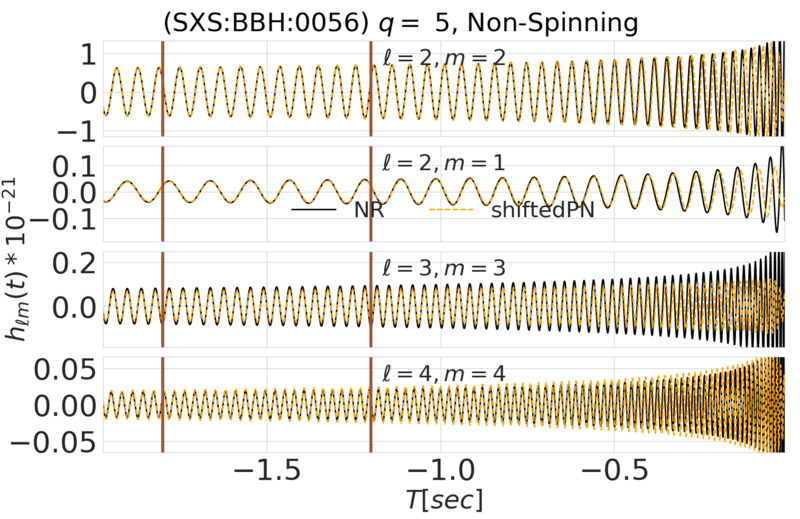}
\includegraphics[width=.85\columnwidth]{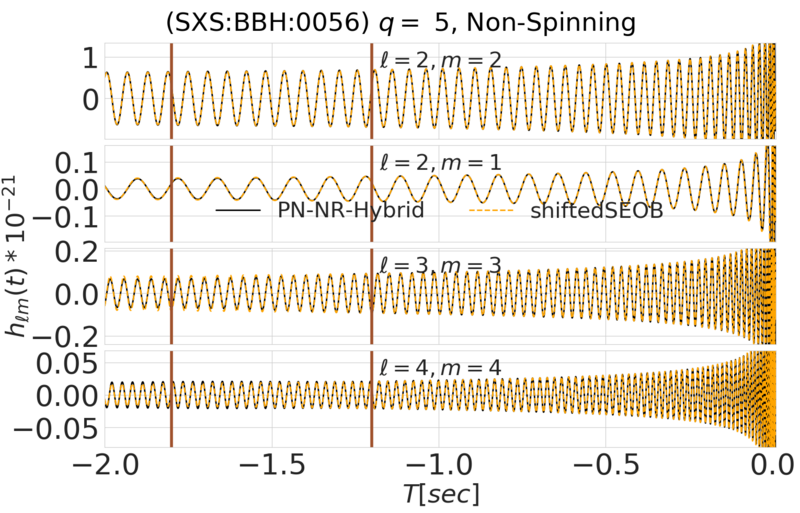}
\includegraphics[width=.85\columnwidth]{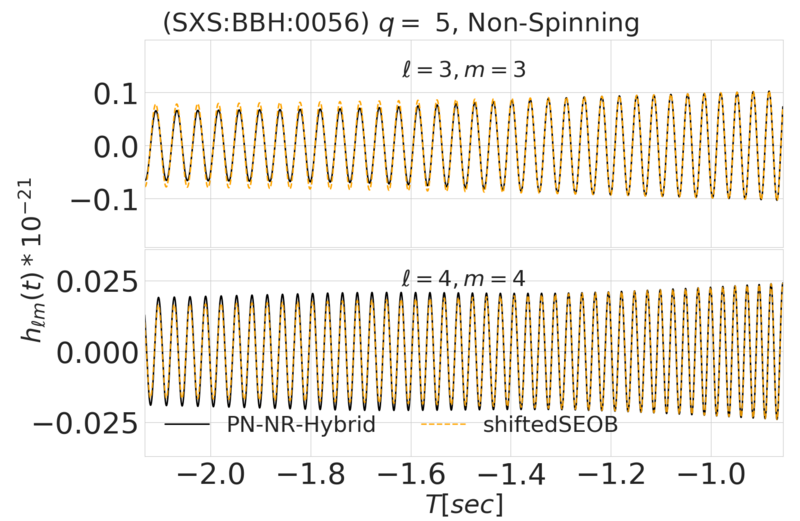}
\includegraphics[width=.85\columnwidth]{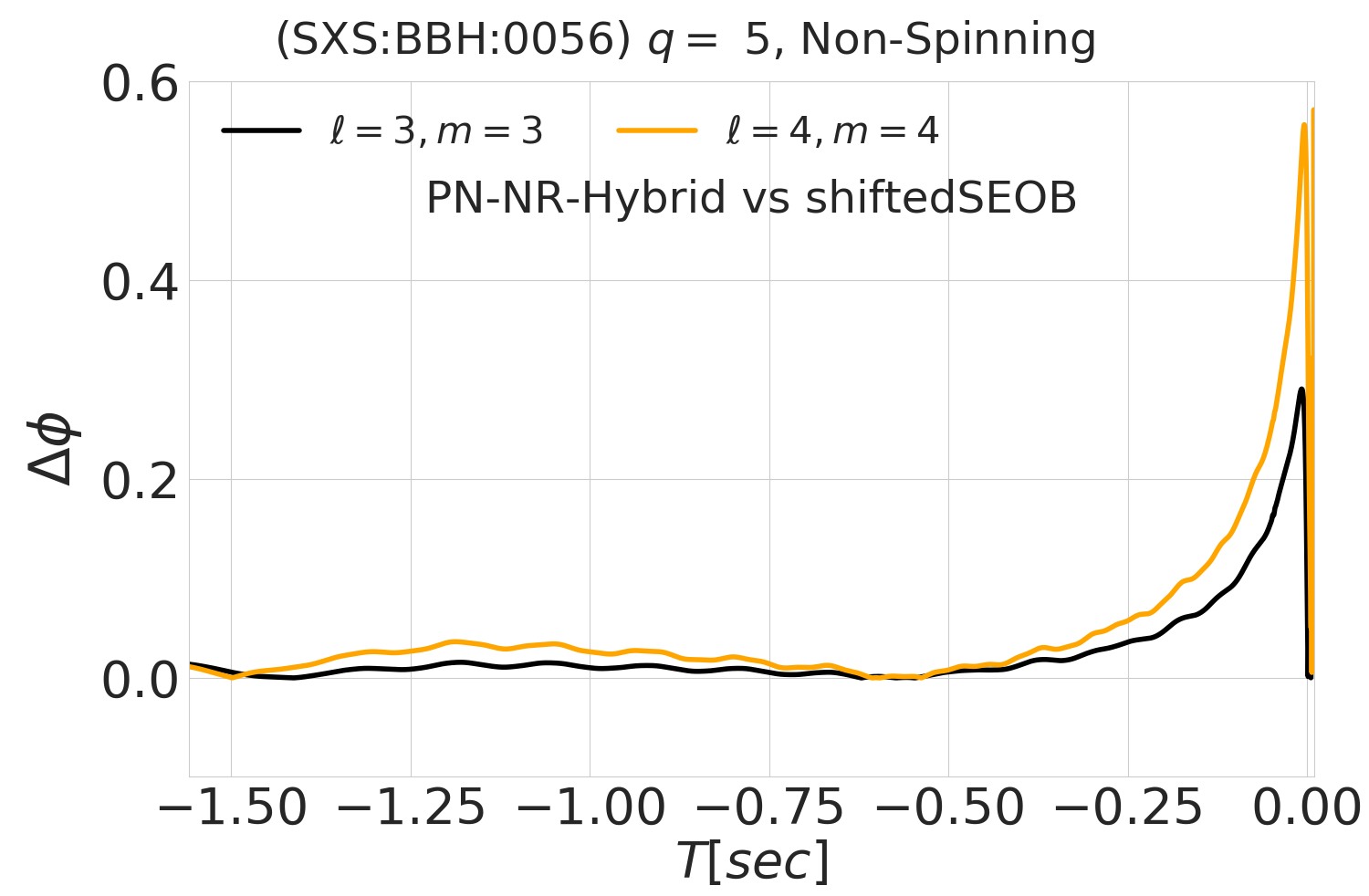}\\
  \caption{Hybridization of a nonspinning, $q=5$ system (SXS:BBH:0056) and a 
 spinning, but nonprecessing case (SXS:BBH:0047)
with $q = 3$ and initial spins $\chi_1= (0,0,0.5)$, $\chi_2= (
0,0,0.5)$ .  
  The numerical waveforms  were obtained  
  from \cite{mroue_abdul_2019_3312192} and  \cite{mroue_abdul_2019_3311869}. The PN waveforms used the spin-Taylor T4 approximant.
  The EOB waveform corresponding to SXS:BBH:0056 and SXS:BBH:0047 was obtained using SEOBNRv4HM.
  The waveforms correspond
  to $M_{\text{tot}} = 70 M_{\odot}$.
  Results from SXS:BBH:0047 are shown on the top row. Results from SXS:BBH:0056 are shown on the bottom two rows.
  The top-left and middle-left plots show the PN and NR modes (after time shifting and phase translations). The top-right and middle-right plots show the hybrid and EOB modes.
 Although not apparent in the plots in the first two rows, there is a nontrivial amplitude error in the $(3,3)$ and $(4,4)$ PN modes. The bottom row shows the hybrid and EOB modes for the $(\ell=3, m=3)$ and $(\ell=4, m=4)$ modes of SXS:BBH:0056. Note the amplitude error in the early part of the waveform. Finally, the plot on the bottom-right shows the phase difference between the hybrid and EOB $(\ell=4, m=4)$ mode for SXS:BBH:0056 near merger.
}

\label{fig:SXS0056ModesPlot}
\label{fig:SXS0047ModesPlot}
\end{figure*}

The resulting hybrid constructed using our method is shown in Fig.~\ref{fig:SXS0056ModesPlot}. The plot shows the NR and PN modes, the resulting hybrid waveforms, and comparisons of the hybrid with the EOB waveform. Note that the $(\ell=4, m=4)$ mode of the PN model has an amplitude error not apparent in EOB mode. We also constructed a hybrid of the NR and EOB waveforms.

   The next case we studied was an aligned spin (and therefore nonprecessing) binary (SXS:BBH:0047).  We constructed two hybrids, one based on the spin-Taylor T4  and NR modes, the other based on the SEOBNRv4HM and NR modes.
The SEOBNRv4HM approximant  has the ($\ell =2, m =\pm 2$), ($\ell=2, m=\pm1$), ($\ell =3, m =\pm 3$), ($\ell =4, m =\pm 4 $), and ($\ell =5, m = \pm 5$) modes. However, we did not use the ($\ell =5, m = \pm5$) modes for our analysis. We show results similar to the nonspinning case in Fig.~\ref{fig:SXS0047ModesPlot}.

\begin{figure*}
\includegraphics[width=.45\textwidth]{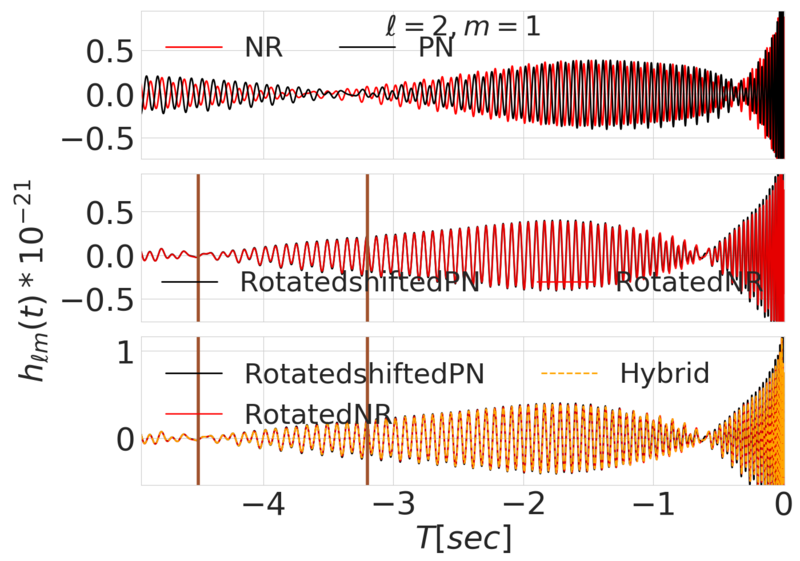}
\includegraphics[width=.45\textwidth]{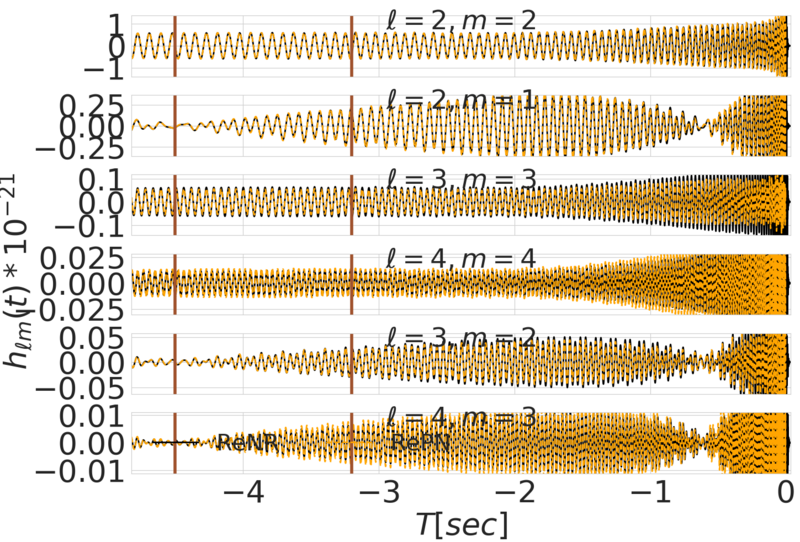}
\includegraphics[width=.45\textwidth]{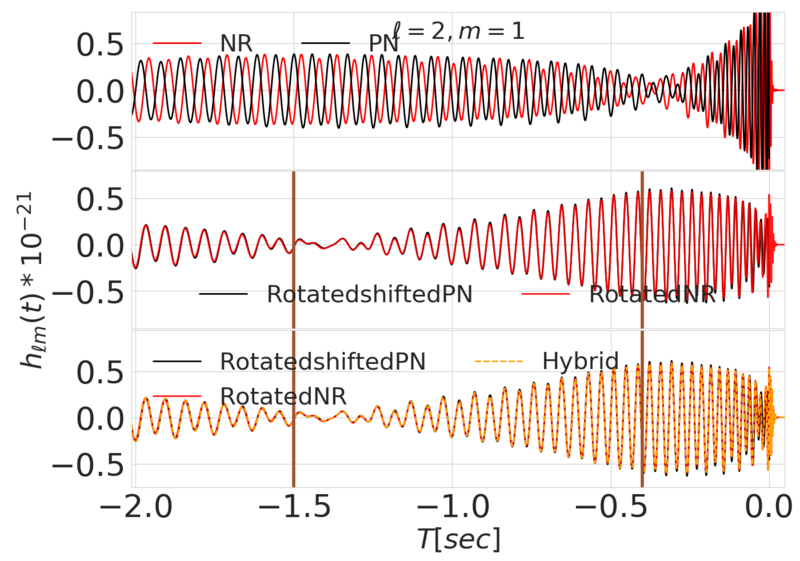}
\includegraphics[width=.45\textwidth]{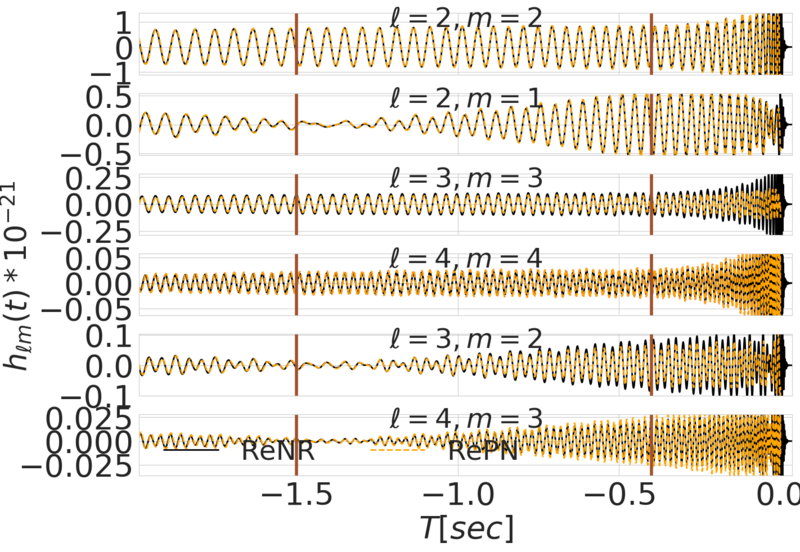}
\includegraphics[width=.45\textwidth]{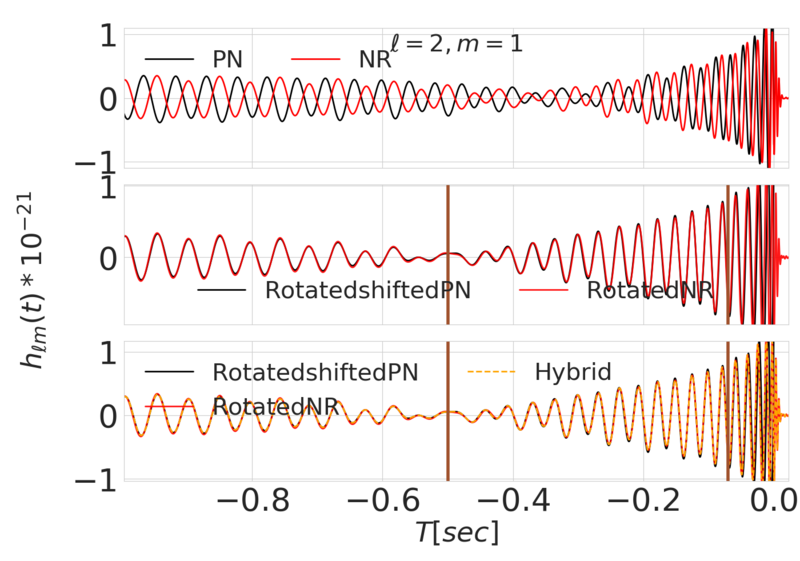}
\includegraphics[width=.45\textwidth]{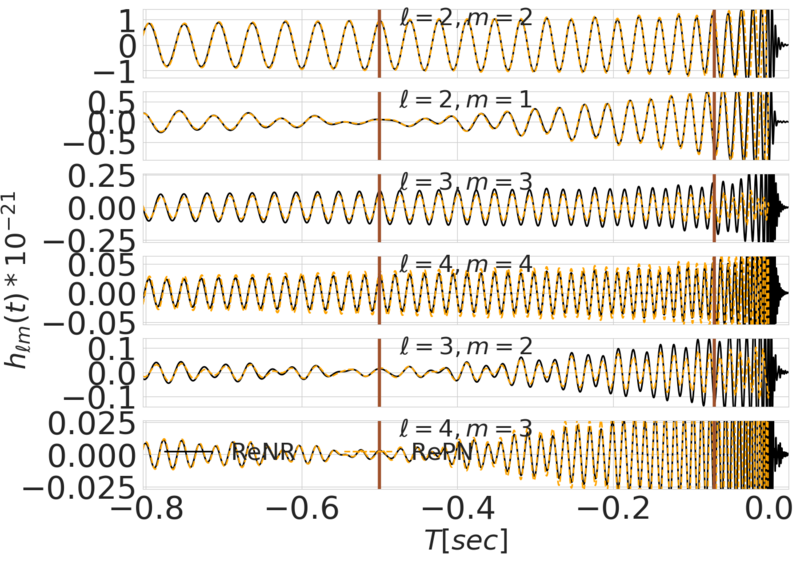}
\caption{ A $q=4$ mildly precessing waveform (BBH1410). The NR and PN waveforms were aligned in the early inspiral regime. 
  The plots on the left show the alignment of the NR and PN $(\ell=2, m=1)$ modes (as well as the hybrid) for the case where the hybridization is done 80 (top), 40 (middle), and 20 (bottom) cycles before merger. For each plot on the left, the top panel shows the ($\ell =2, m=1$) modes before alignment, the middle panel shows the modes after alignment, and the bottom panel shows the hybrid overploted onto the aligned modes. 
  The plots on the right show the rest of the PN and NR modes after alignment for these three cases. The waveforms correspond to $M_{\text{tot}} = 70 M_{\odot}$. In all cases, the vertical lines show the hybrid interval.}
\label{fig:SXS1410EarlyHybrid}
\label{fig:SXS1410LateHybrid}
\label{fig:SXS141020CyclesHybrid}

\end{figure*}

We next consider three mildly precessing cases. Our goal here was to use  very long numerical waveforms and then truncate
them. We then compare the hybrids of the truncated waveforms with the original numerical waveforms.  The first case we
considered is the  SXS1410 waveform  \cite{sxs_collaboration_2019_3315705} (see Table~\ref{tab:AllwaveformUsed}).  We
used the spin-Taylor T4 approximant for post-Newtonian waveforms based on \cite{Ajith:2011ec} and obtained from
\cite{lalsuite}. We choose the initial frequency for PN waveforms to be the same as the initial frequency of the numerical waveform
(prior to truncating the waveform). We choose $f_{\textrm{ref}} = 8.5045 {\rm Hz}$ which was approximately the initial frequency of the numerical waveform (recall that the hybrid is constructed with a binary mass of $70 M_\odot$ and then rescaled to different masses). The spin configurations were chosen to be the same as the initial spin configurations of the numerical waveforms. We choose $\phi_{\textrm{ref}}$ to be zero, which means the large black hole (BH) is at along the $x$-axis initially. First, we hybridized the two waveform earlier in inspiral regime. This corresponds to $80$ cycles before
merger. We then hybridized them closer to merger 40 cycles before merger. Finally, we hybridized waveforms 20 cycles
before merger. The resulting aligned waveforms are shown in Fig.~\ref{fig:SXS1410EarlyHybrid}.

In addition, we considered two other mildly precessing waveforms. We used
the SXS1392 simulation \cite{sxs_collaboration_2019_3315664} as well the RIT  simulation RIT0137 \cite{Healy:2019jyf} (see Table~\ref{tab:AllwaveformUsed}).
 Again we hybridized them with spin-Taylor T4 PN waveforms in early inspiral, as well as late inspiral phase.  The PN
 waveform is obtained by setting initial frequency to be same as numerical waveforms which in this case was
 $f_{\textrm{ref}} = 4.73801 {\rm Hz}$ and $f_{\textrm{ref}} = 11.6455) {\rm Hz}$, respectively (at $M=70 M_\odot$).

\section{Analysis}
\label{sec:analysis}

To asses the accuracy and usefulness of our hybridization procedure we calculate the mismatch between the hybrid waveform and either very long NR waveforms, or model waveforms (e.g., EOB). The mismatch itself is calculated in two ways. First, we perform a mode-by-mode mismatch using the \textit{CreateCompatibleComplexOverlap} function in {\sc LaLSimUtils}. This function automatically optimizes over both time translations and phase shifts. Because of this, the mode-by-mode mismatch allows for the phase shifts of different modes to be inconsistent. That is, one expects each $m$ mode to be shifted by $m\phi$. As a second analysis, we construct a grid of angles that covers the sphere and calculate the mismatch at each point on the grid. We then plot the results. This latter analysis guarantees that all modes are time and phase shifted consistently, but suffers from the fact that the $(\ell=2, m=\pm2)$ modes will dominate the mismatch calculation.

First, we define an inner product
\begin{equation} 
   \left \langle h_1| h_2 \right \rangle = 2 \,  \int_{-\infty}^{\infty} \frac{h_1^*(f) h_2(f)}{S_n(f)}  
\mathrm{df}
\end{equation} 
where  $h(f)$ is the Fourier transform of the complex waveform $h(t)$ and we use the Advanced-LIGO       
design sensitivity \emph{Zero-Detuned-HighP} noise curve \cite{ligonoisecurve} with $f_{\text{min}} =$   
20Hz and  $f_{\text{max}} =$ 2000Hz. This inner product can also be computed with a further maximization 
over time and phase shifts as described in \cite{Ohme:2011zm},
\begin{equation} 
   \left \langle h_1| h_2 \right \rangle = \mathop{max}_{t_0,\phi_0}\bigg[ 2\,\left |  \int_{-           
\infty}^{\infty} \frac{h_1^*(f) h_2(f)}{S_n(f)} \mathrm{df} \right |  \bigg]
\end{equation}
The overlap of two waveforms is then given by
\begin{equation} 
  \mathcal{O}=    \frac{   \left \langle h_1| h_2 \right \rangle }{\sqrt{   \left \langle h_1| h_1       
\right \rangle  \left \langle h_2| h_2 \right \rangle  }} 
\end{equation}
and the mismatch is  given by
\begin{equation} \label{eq:mm_int}
  \mathcal{M} = 1 - \mathcal{O}
  \end{equation}
The mismatch indicates how close the two waveforms $h_1$ and $h_2$ are, with a mismatch of 0 indicating  
the two waveforms are essentially the same. If $\mathcal M$ is less than some threshold, we regard the   
final hybrid as \emph{accurate enough} for detections. For a maximum loss of 10\% of the signals in the  
detection process, we can accept a mismatch of no more than 1.5 \% \cite{Hannam:2010ky} or even 0.5\%,   
as suggested in \cite{Lindblom:2008cm}.

\begin{figure*}
\includegraphics[width=.47\textwidth]{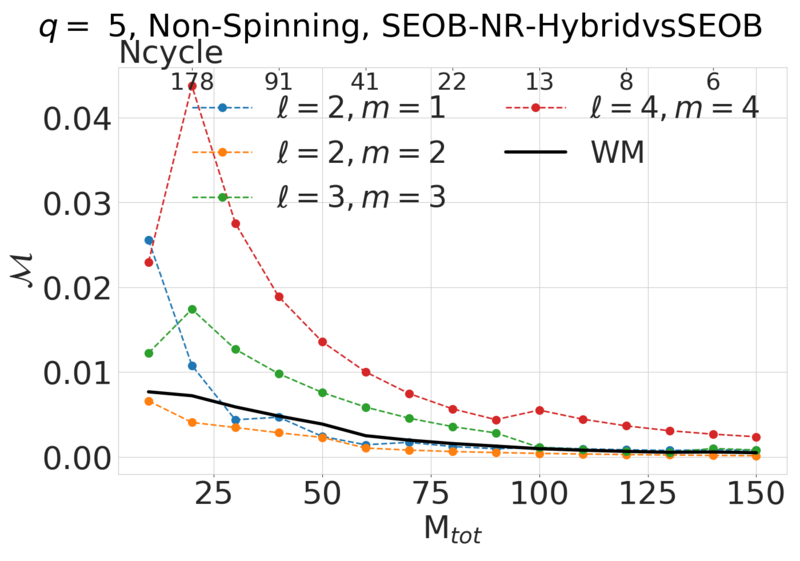}
\includegraphics[width=.47\textwidth]{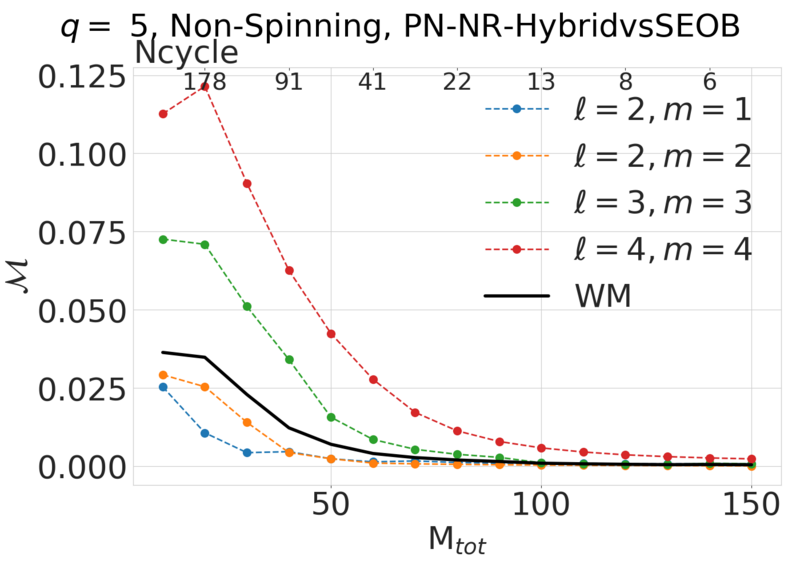}
\includegraphics[width=.47\textwidth]{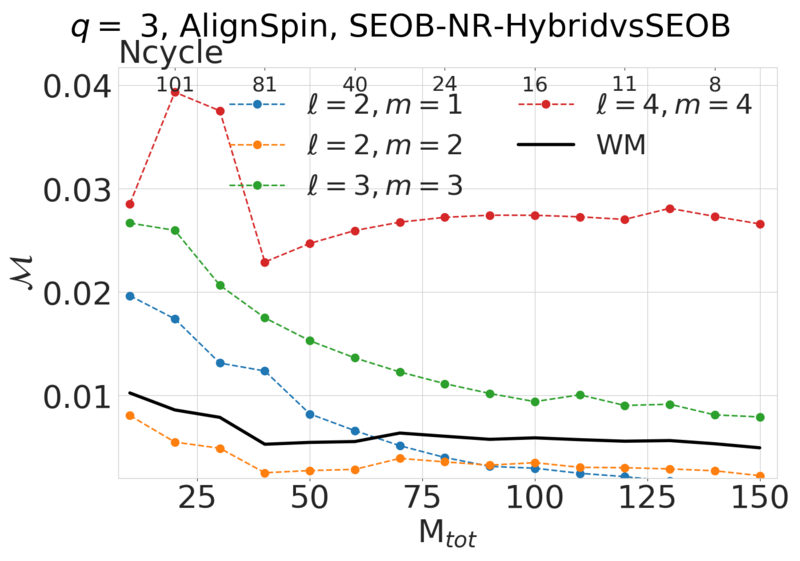}
\includegraphics[width=.47\textwidth]{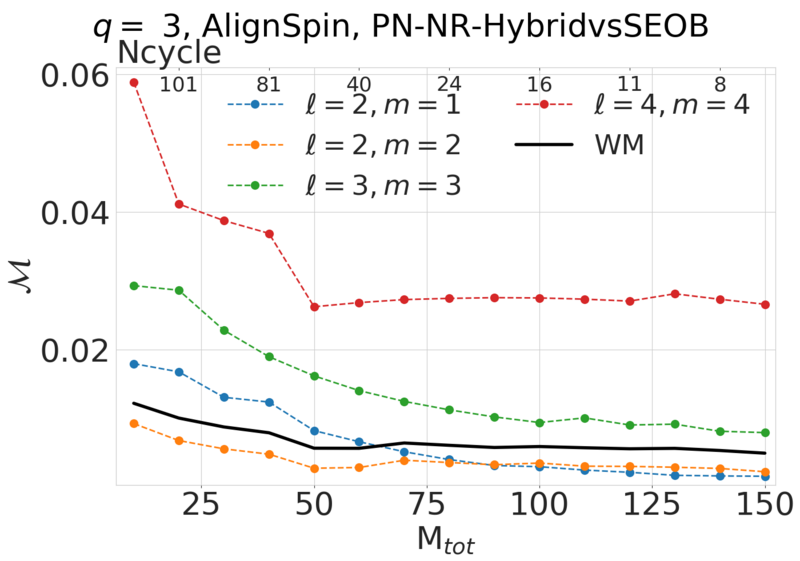}
  \caption{
    Mismatch for different modes for the two nonprecessing cases. The top two plots  are for the  SXS:BBH:0056 (nonspinning, $q=5$) configuration and the bottom two plots  are for the SXS:BBH:0047 ($q=3$, aligned spins). The numerical waveforms were obtained from \cite{mroue_abdul_2019_3312192} and corresponding post-Newtonian and EOB waveform are taken from \cite{lalsuite}. The plots show the EOB-NR hybrid versus the EOB waveform and the PN-NR hybrid versus the EOB waveform.
  The largest errors are in the $(\ell=4, m=4)$ mode.  We use the Advanced-LIGO design sensitivity \emph{Zero-Detuned-HighP} noise curve \cite{ligonoisecurve} with $f_{\text{min}} =$ 20Hz and  $f_{\text{max}} =$ 2000Hz.  Finally, the curves marked WM are the weighted mismatch defined by Eq.~(\ref{eq:WM}).
 On each plot, the top axis shows the number of cycles to merger with frequencies larger than 20 Hz (which is a function of the total mass). Here, WM refers to the mode-weighted mismatch. }
\label{fig:Mismatchnonprecess}
\end{figure*}

We begin our analysis by comparing the hybrid of the nonspinning waveform (SXS:BBH:0056) to the corresponding EOB waveform. As explained above, we computed two different hybrids: an NR-EOB hybrid and an NR-PN hybrid. The mode-by-mode mismatch versus the total mass of the binary is given in Fig.~\ref{fig:Mismatchnonprecess}. At early times, the PN and EOB waveforms disagree substantially in the $(\ell=4, m=4)$ and $(\ell=3, m=3)$ modes, which is apparent in the mismatch between the PN-NR and EOB waveforms at small masses. On the other hand, the $(\ell=4, m=4)$ mode of the EOB-NR and EOB waveforms disagree by more than $1.5\%$ at high masses. This, in turn means that EOB $(\ell=4, m=4)$ mode, as shown in Fig.~\ref{fig:SXS0056ModesPlot}, has a relatively large phase difference to the NR mode when compared to the lower-order modes.  
We see similar behavior for the spinning, but nonprecessing system (SXS:BBH:0047).  The mismatch between the PN and EOB $(\ell=3, m=3)$ modes and $(\ell=4, m=4)$ modes is larger than our cutoff tolerance of 1\% at all masses. On the other hand, we see that the EOB and NR waveforms for the $(\ell=4,m=4)$ modes show a mismatch of $2.5\%$ (as is evident by the high-mass limit in the plots). This indicates a significant offset of the EOB version of this mode from the numerical one.

While the mode-by-mode mismatch measures the errors in each mode, it accounts for neither the relative power in each mode nor the orientation-dependence of the
mismatch.  For example, motivated by the orientation-averaged overlap $\int \frac{d\Omega}{4\pi} \qmstateproduct{h_1}{h_2} =
\sum_{lm} \qmstateproduct{h_{1,lm}}{h_{2,lm}}/4\pi$, we can introduce a mode-weighted mismatch
\begin{eqnarray}
W[{\cal M}] = \frac{\sum_{lm} \rho_{lm}^2 {\cal M}_{lm}}{\sum_{lm} \rho_{lm}^2}
  \label{eq:WM}
\end{eqnarray}
where ${\cal M}_{lm}$ are the mode-by-mode, time-and-phase-maximized mismatches and $\rho_{lm}^2=\qmstateproduct{h_{lm}}{h_{lm}}$.   The dark black curve on each of the mismatch figures shows the corresponding mode-weighted mismatch. For the nonprecessing case, this weighted mismatch closely follows the dominant quadrapolar mismatch curves. As we will see below, nonquadrapolar mismatches become increasingly important in the precessing case.

For the precessing case, we do not have models whose systematic errors are confidently well below the hybridization
errors we seek to assess.  Rather, we compare the hybrid waveform with a much longer numerical waveform, as explained above. One consequence of this choice is that at high masses, the model waveform (i.e., the original NR waveform) and the hybrid are essentially identical.

\begin{figure*}
\includegraphics[width=.41\textwidth]{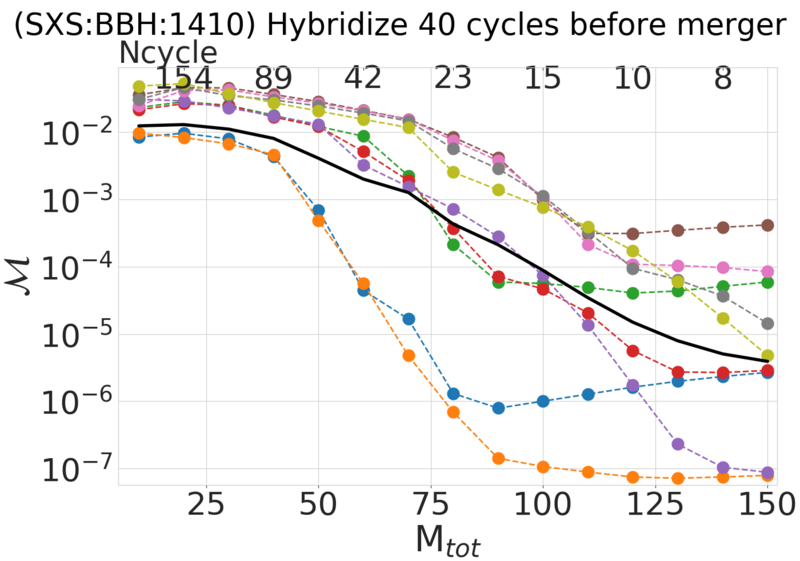}
\includegraphics[width=.41\textwidth]{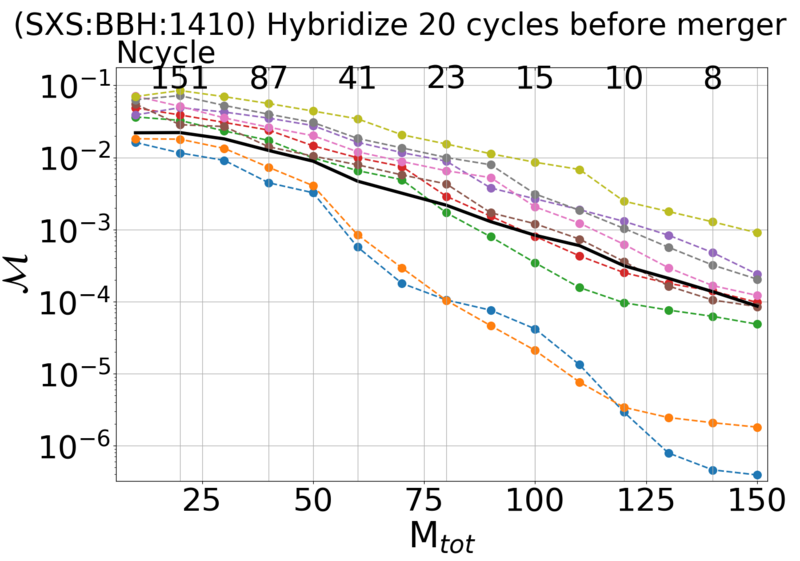}
\includegraphics[width=0.41\textwidth]{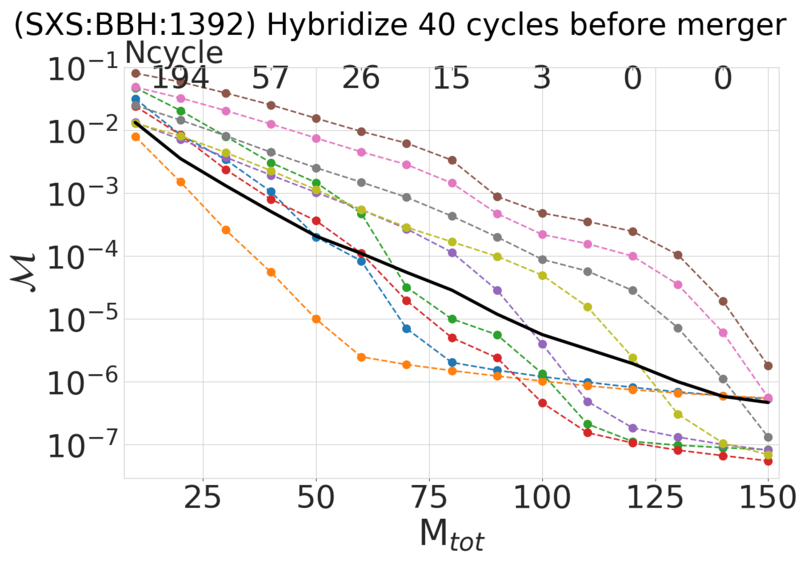}
\includegraphics[width=0.41\textwidth]{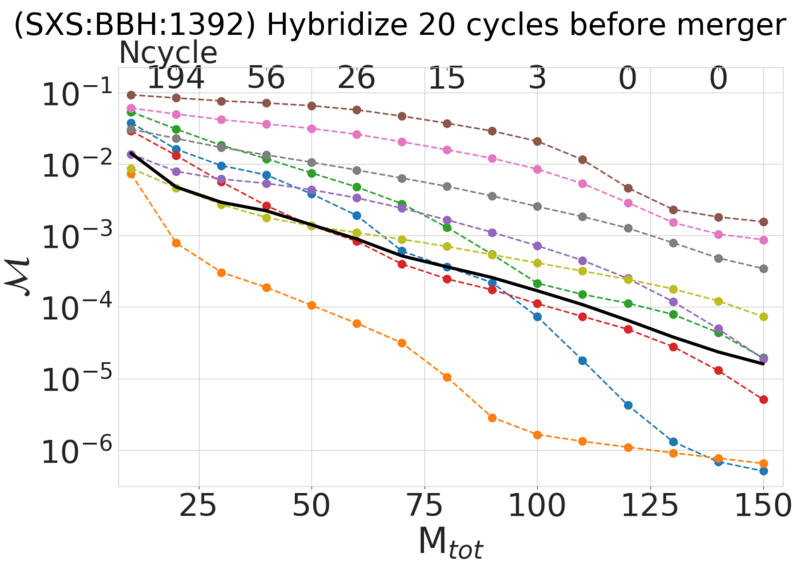}
\includegraphics[width=.41\textwidth]{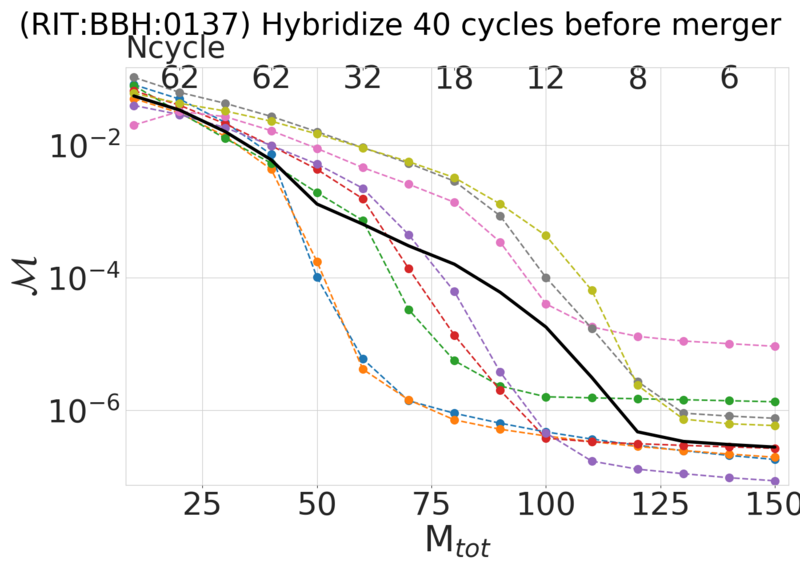}
\includegraphics[width=.41\textwidth]{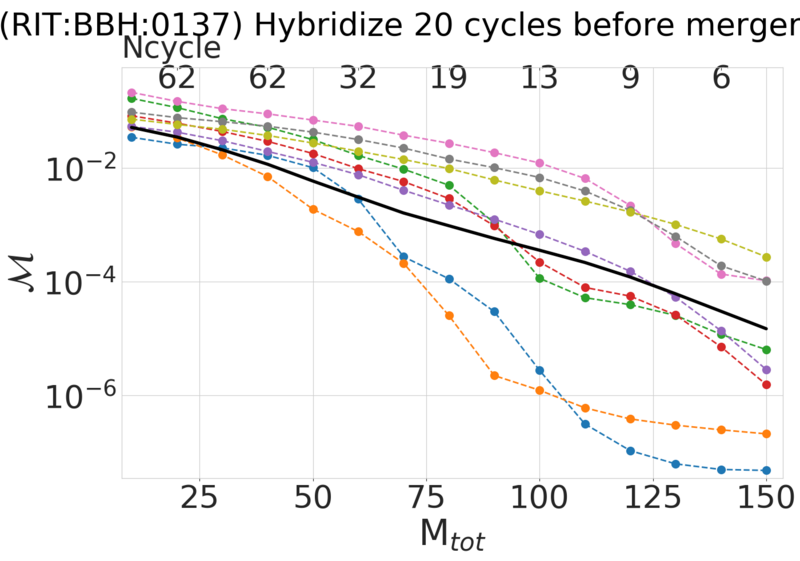}
\includegraphics[width=.16\textwidth]{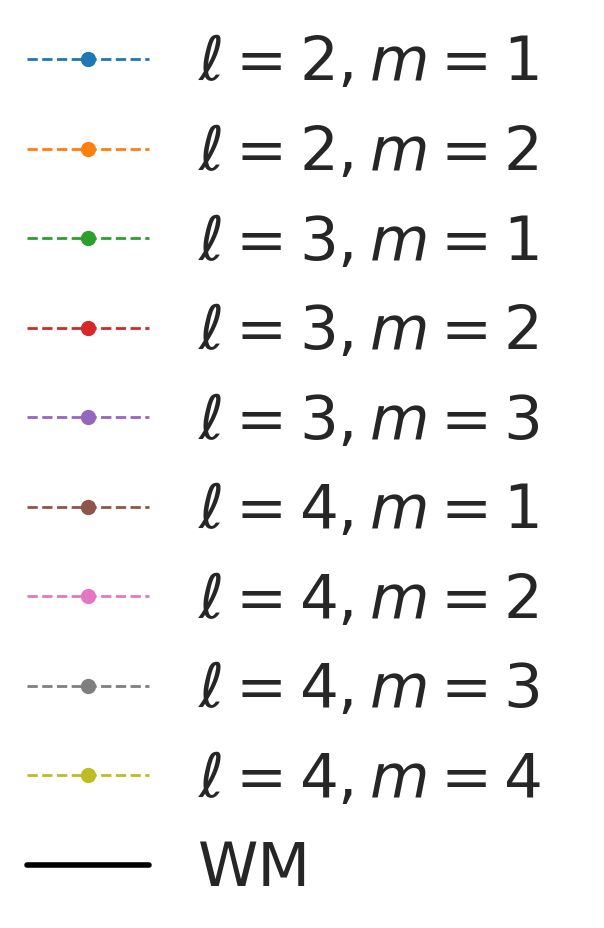}
  \caption{Mismatch as a function of total mass for different $(\ell, m)$ modes for the three precessing systems. The top panels show the mismatched for
 SXS:BBH:1410. The second row shows the mismatched for SXS:BBH:1392, and the third row shows the mismatches for RIT:0137.
  The numerical waveforms were taken from \cite{sxs_collaboration_2019_3315705} and \cite{RIT:catalog}, respectively, and the post-Newtonian waveforms taken from \cite{lalsuite} based on \cite{Ajith:2011ec}. Hybridization is done in both inspiral as well as late closer to merger regions. The plots  show the result when the hybrid is constructed  40 cycles before merger and  20 cycles prior to merger. We use the Advanced-LIGO design sensitivity \emph{Zero-Detuned-HighP} noise curve \cite{ligonoisecurve} with $f_{\text{min}} =$ 20Hz and  $f_{\text{max}} =$ 2000Hz.   The curves marked WM are the weighted mismatch defined by Eq.~(\ref{eq:WM}). On each plot, the top axis shows the number of cycles to merger with frequencies larger than 20 Hz (which is a function of the total mass).}
\label{fig:AllmodesmismatchearlyinspiralSXS1410}
\label{fig:AllmodesmismatchEarlyinspiralSXS1392}
\end{figure*}

We show the mode-by-mode mismatches for the three precessing cases in Fig.~\ref{fig:AllmodesmismatchearlyinspiralSXS1410}.
In the figure, we show the mismatch between two hybrids and the original NR waveforms. One of these hybrids is constructed starting 40 cycles and the other at 20 cycles before the merger. 
For the former case,  the higher-order modes fall within the $1\%$ tolerance for masses larger than $60 M_\odot$ and $80 M_\odot$, for the $\ell=3$ and $\ell=4$ modes, respectively. For the hybrid constructed 20 cycles prior to merger, the $\ell=4$ mismatched are within tolerance at $95 M_\odot$.
The mismatch at small masses indicates a substantial phase difference between the PN modes used to construct the hybrid and the early part of the numerical waveform (note, the hybrid is constructed from the late part of the NR waveform). In addition, we include  the weighted mismatches as a function of  total mass and number of cycles in the numerical waveform in Table~\ref{tab:IntervalMismatch}.
\begin{figure*}
\includegraphics[width=.41\textwidth]{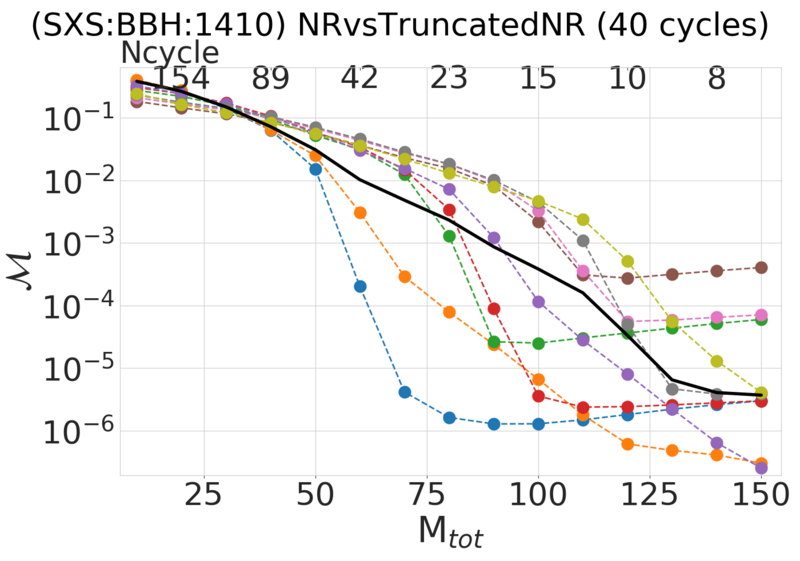}
\includegraphics[width=.41\textwidth]{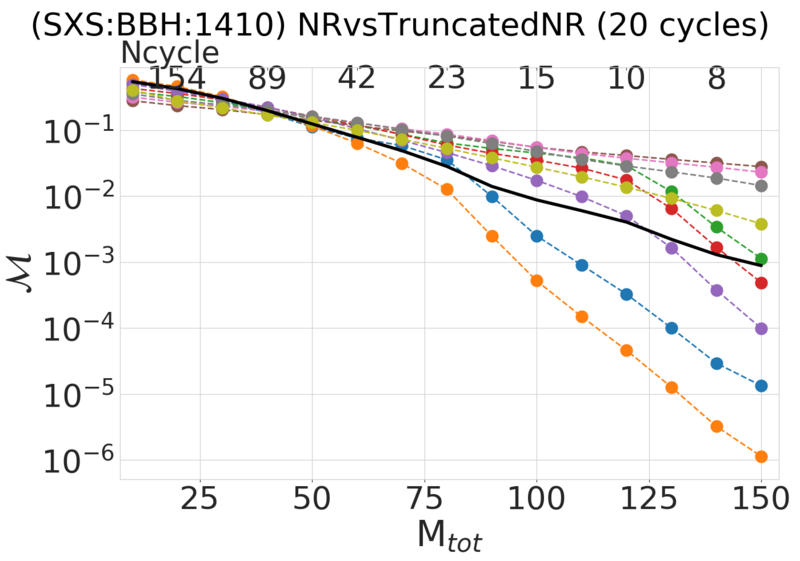}
\includegraphics[width=.41\textwidth]{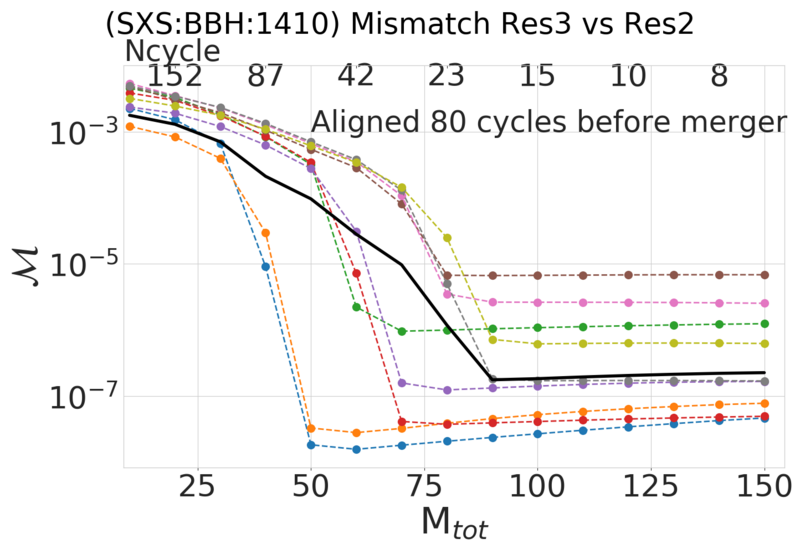}
\includegraphics[width=.41\textwidth]{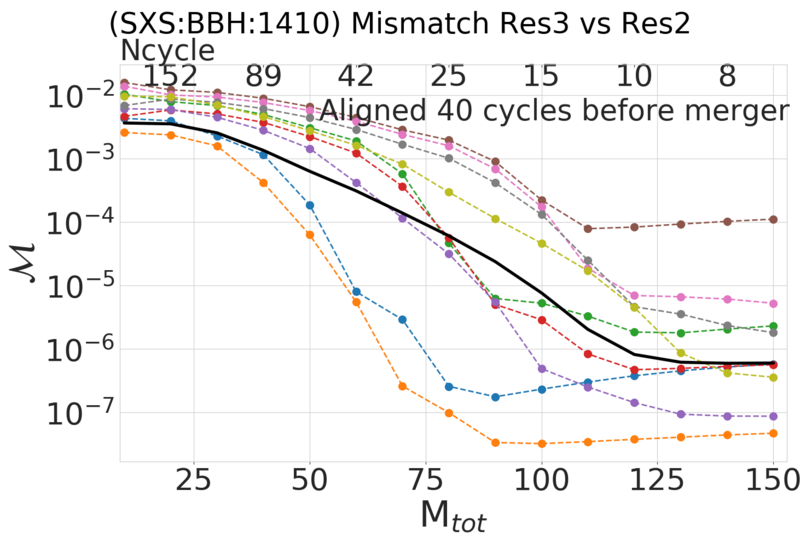}
\includegraphics[width=.16\textwidth]{legend.png}
  \caption{(top panels) Plots of the mismatch between the full NR waveform against the NR waveforms truncated at 40 and 20 cycles prior to merger (i.e., truncated but not hybridized)
  for SXS:BBH:1410. Note the factor of $\sim10$ improvement in the mismatch when hybridizing the waveform, as shown in Fig.~\ref{fig:AllmodesmismatchearlyinspiralSXS1410}. (bottom panels) Plots of the mismatch between the Lev2 (low resolution) and Lev3 (high resolution) waveforms for SXS:BBH:1410. Here the mismatch is between the two NR waveforms.}
\label{fig:MismatchRes2vsRes3}
\end{figure*}

One important question that we need to address is to what extent is the mismatch observed is an artifact of numerical
truncation error. To test this, we compute the mismatch of the Lev2 and Lev3(higher resolution) waveforms of SXS1410
(i.e, the two highest resolutions). We calculate the mismatch between the numerical waveforms at these two resolutions 
after aligning the waveforms at 80 cycles prior to merger and again at 40 cycles prior to merger. The results are shown in Fig.~\ref{fig:MismatchRes2vsRes3}.  When aligning the waveforms at 80 cycles, the mismatch is within tolerance for all modes and all masses. On the other hand,  when aligning the waveform at 40 cycles, the mismatches are below tolerance for all modes when the mass is larger than $40 M_\odot$.  Importantly, these mismatches are below those observed for the hybrid. 

Finally, we address the issue of the efficacy of hybridization in the first two plots of the bottom row of Fig.~\ref{fig:MismatchRes2vsRes3}. Here, we plot the mismatch of the original NR waveform with truncated versions of the same waveform. Here, we truncate at 40 and 20 cycles prior to merger. 
When we truncate the waveform at 40 cycles before merger, the weighted mismatch is outside the tolerance of $1\%$ for $M < 60 M_\odot$, while the corresponding hybrid is within tolerance for $M> 40 M_\odot$. The improvement is more dramatic for the 20 cycles case. When truncating at 20 cycles, even the $\ell=2$ modes are outside tolerance for total masses less than 80 $M_\odot$, and the weighted mismatch is outside the $1\%$ tolerance for $M< 100 M_\odot$.  The corresponding hybridized waveform is within tolerance for all masses larger than $50M_\odot$.

\begin{figure}
\includegraphics[width=.95\columnwidth]{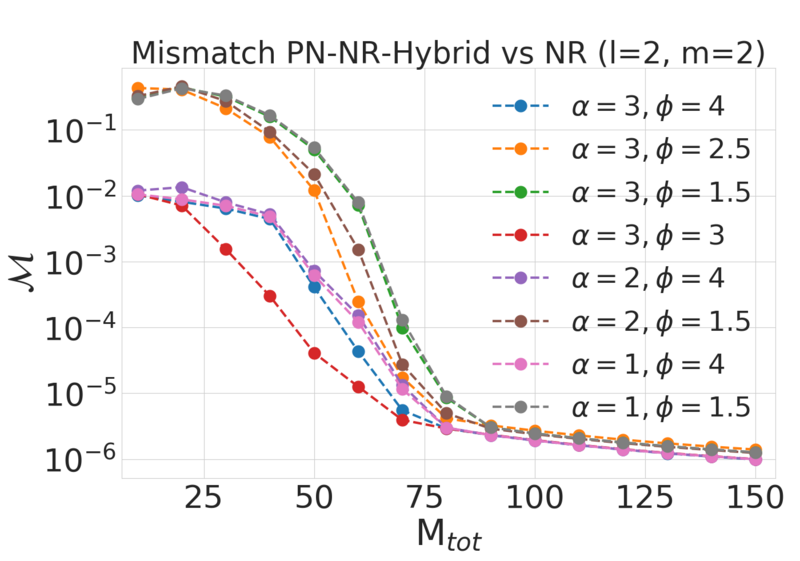}
\includegraphics[width=.95\columnwidth]{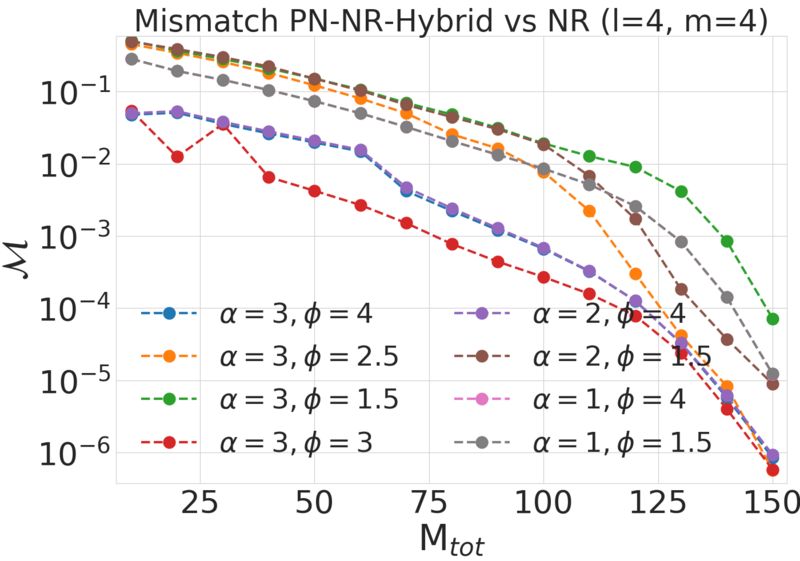}
  \caption{Mismatch of the $(\ell=2, m=2)$ and $(\ell=4, m=4)$ modes of the PN-NR hybrid (versus NR) for the SXS:BBH:1410 case versus PN order. Here, $\alpha$ is the amplitude order and $\phi$ is the phase order.
  In general higher order approximants will provide more accurate hybridization although the phase order or 3 provides the most accurate hybrid.}
\label{fig:PNorderterms}
\end{figure}

Thus far, we have considered how truncation errors in the NR waveforms can affect the mismatch (see Fig.~\ref{fig:MismatchRes2vsRes3}). In Fig.~\ref{fig:PNorderterms}, we consider how PN truncation errors affect the accuracy of the resulting hybrid. To do this, we modify the amplitude order (here denoted by $\alpha$) and the phase order (here denoted by $\phi$) of the PN approximation.
We use the  spin-Taylor T4 approximant in all cases and find that, in general, higher PN order terms in both amplitude and phase lead to more accurate hybridization although, we found the best result was obtained with 3PN order terms in the phase rather than the 3.5 or 4th order terms. For a detailed analysis of PN errors in the waveform see~\cite{Ohme:2011zm}.  We see that the PN truncation error has a substantial effect on the mismatch, which indicates that the PN truncation error is the dominant error at the separations considered here.

As an alternative analysis of the mismatch presented is above, for each simulation, we can directly compute the mismatch ${\cal M}$ between the original NR simulation
and our PN-NR hybrid as a function of angle.  Just like the  mismatches in Fig. \ref{fig:MismatchRes2vsRes3}, our choice of
fiducial mass has a significant impact on the overall scale of the mismatch.

\begin{figure*}
\includegraphics[width=.23\textwidth]{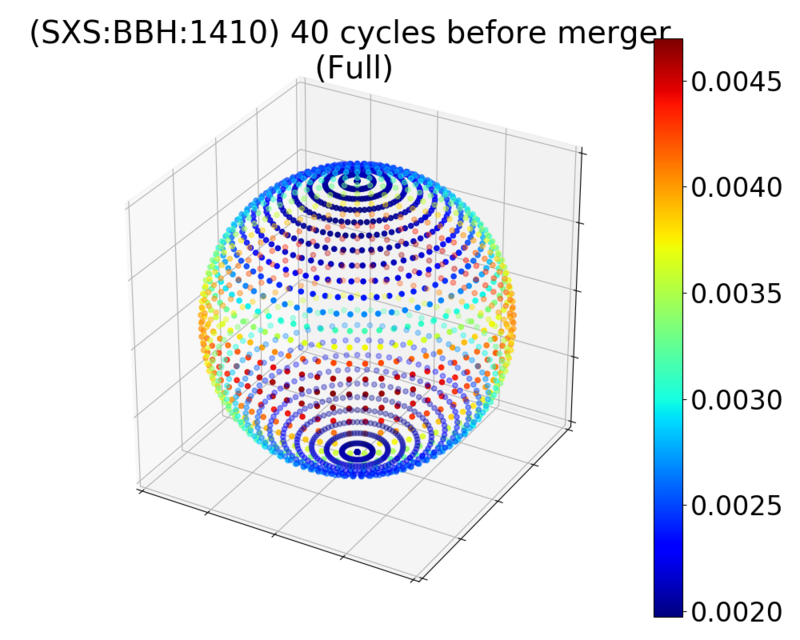}
\includegraphics[width=.23\textwidth]{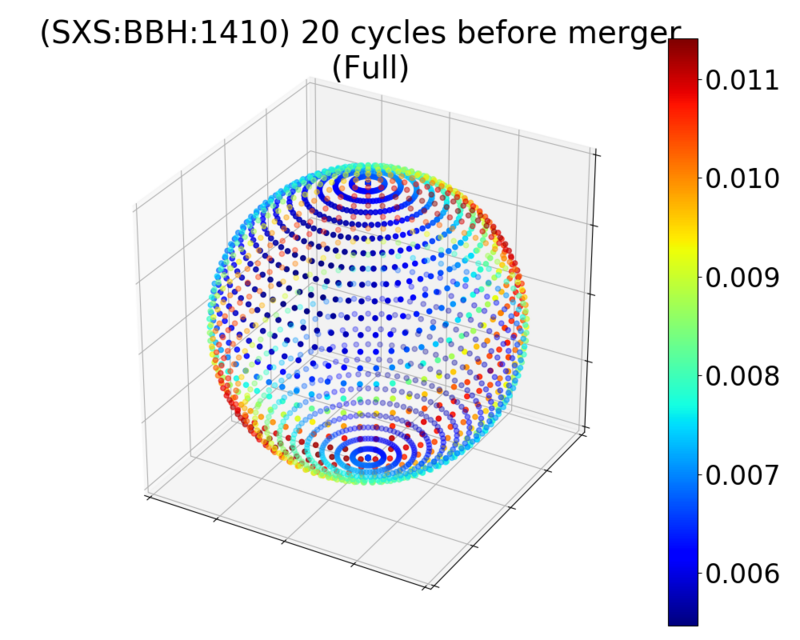}
\includegraphics[width=.23\textwidth]{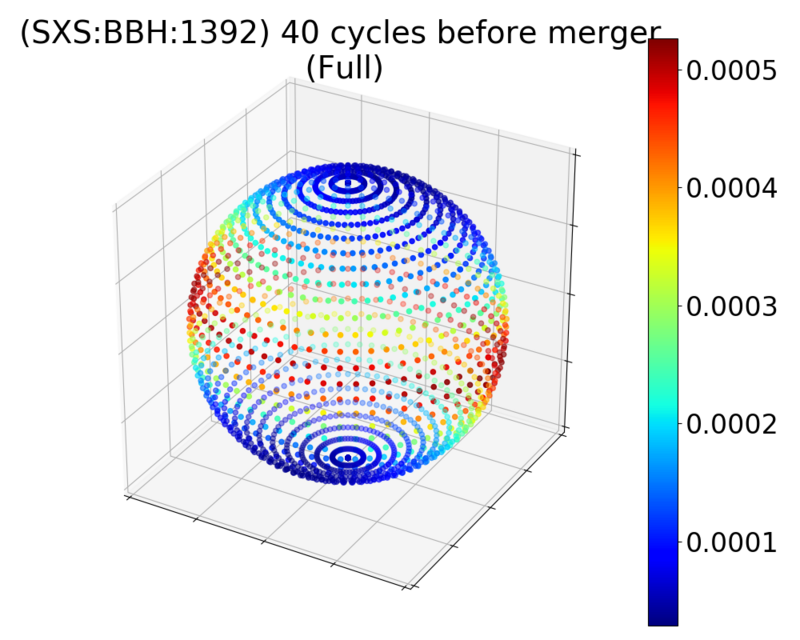}
\includegraphics[width=.23\textwidth]{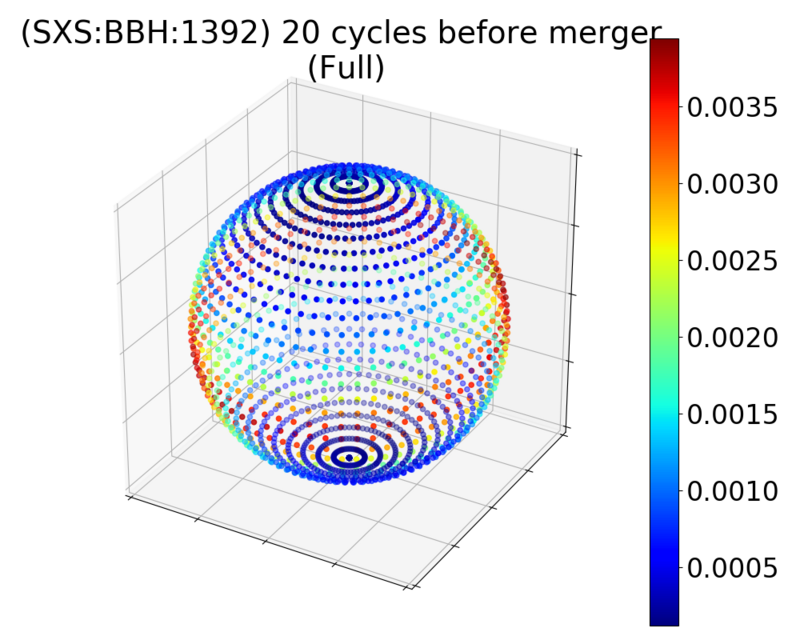}

\includegraphics[width=.23\textwidth]{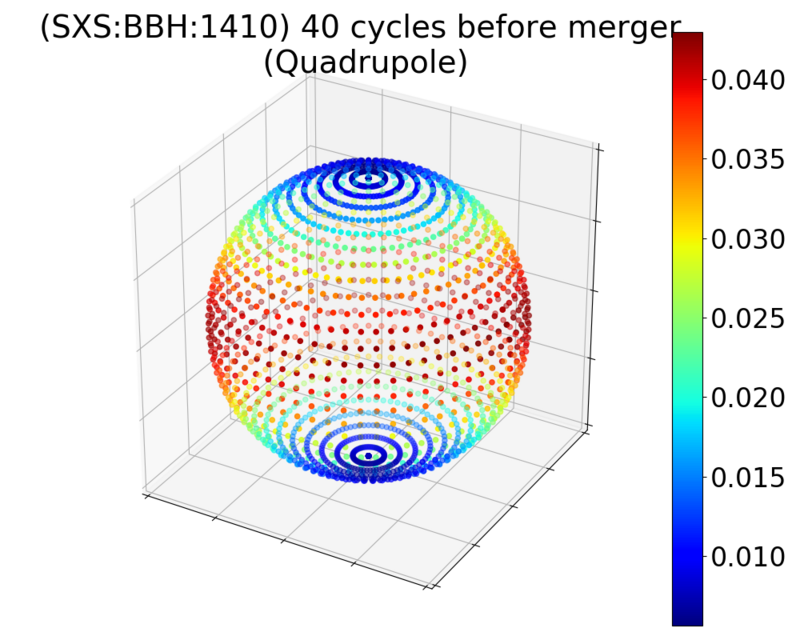}
\includegraphics[width=.23\textwidth]{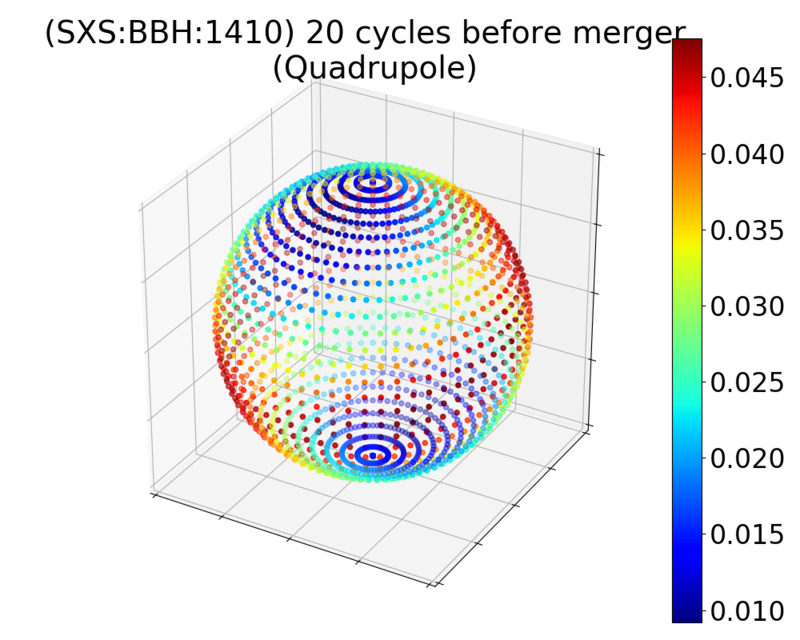}
\includegraphics[width=.23\textwidth]{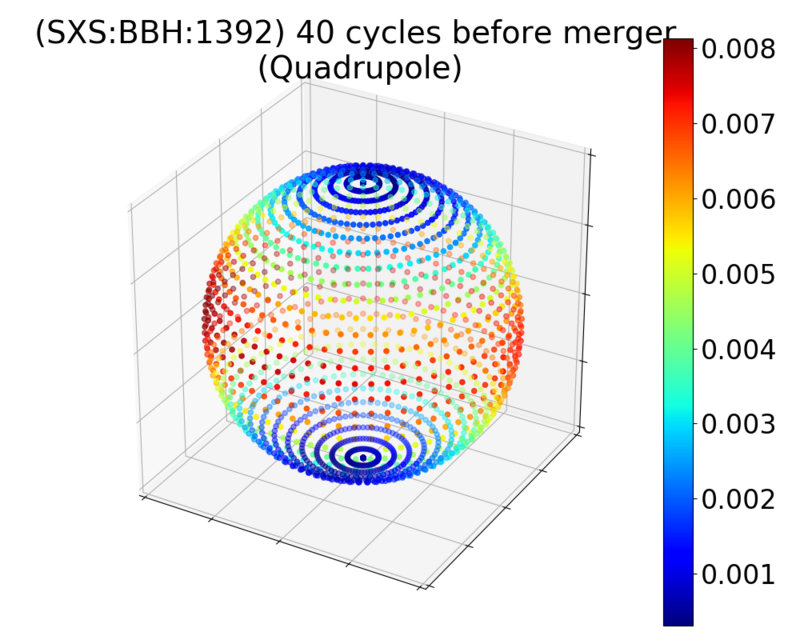}
\includegraphics[width=.23\textwidth]{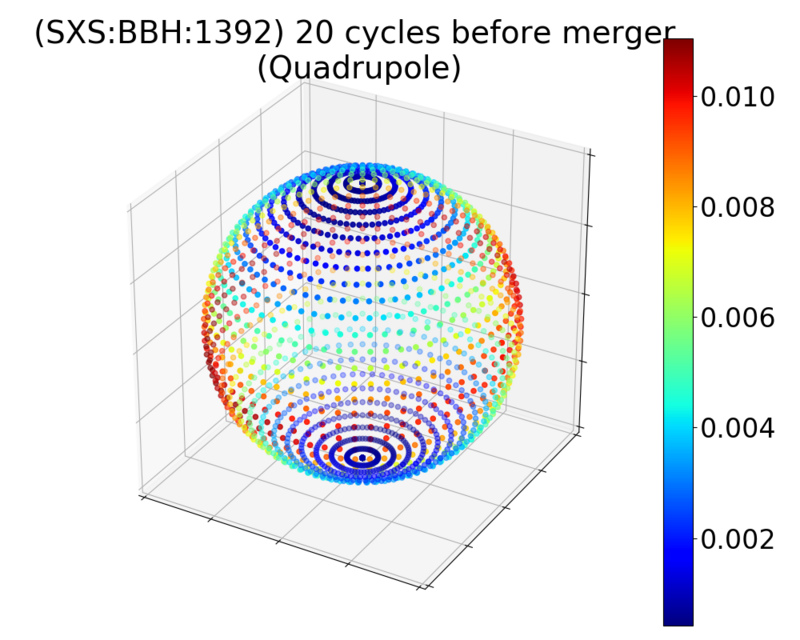}
\caption{ Mismatch of the strain constructed using all hybrid modes and the numerical waveforms for the precessing cases
  SXS:BBH:1410 (left two panels) and  SXS:BBH1392 (right two panels) 
for binaries with a total mass of $40M_{\odot}$. The numerical waveform is taken from \cite{sxs_collaboration_2019_3315705} and post-Newtonian waveforms taken from \cite{lalsuite} based on \cite{Ajith:2011ec}. 
  The plots labeled as {\it Quadrapole} included only the $\ell=2$ modes in the hybrid and comparing to the full NR waveforms (all modes). 
   We use the Advanced-LIGO design sensitivity \emph{Zero-Detuned-HighP} noise curve \cite{ligonoisecurve} with $f_{\text{min}} =$ 20Hz and  $f_{\text{max}} =$ 2000Hz. }
\label{fig:StrainEarlyInspiralSXS1410}
\label{fig:StrainEarlyInspiralSXS1392}
\end{figure*}

In Fig.~\ref{fig:StrainEarlyInspiralSXS1410}, we show the mismatch as a function of angle for a total mass of $40
M_\odot$ for the three precessing simulations. 
 To quantify  the effect that higher-order modes have on the mismatch, we suppress these modes in the hybrid. 
For the two SXS simulations, high-order modes are very important to the total mismatch, with the mismatch increasing by a factor of $\sim 10$ when these modes are suppressed. One the other hand, in the RIT simulation, the quadrapolar modes dominate the waveform. The reason for this difference in behavior between the RIT and SXS waveforms appears to be due to the degree with which the various simulations precess. 

\begin{table*}[t]
\centering
\begin{tabular}{|l|c|c|c|c|c|c|}
\hline
	\multicolumn{7}{|c|}{(SXS:BBH:1410) Mismatch versus Frequency of Hybrid} \\
\hline
\multicolumn{1}{|c|}{Cycles $\Longrightarrow$ } & \multicolumn{2}{c|}{10} & \multicolumn{2}{c|}{20}  & \multicolumn{2}{c|}{40}\\
\cline{2-7}
	\multicolumn{1}{|c|}{$M_{tot}\; \Downarrow $} & Freq(Hz) & $\mathcal{M}$ & Freq (Hz)& $\mathcal{M}$   & Freq(Hz) & $\mathcal{M}$ \\
\hline
10   &  [210.7-341.5] & 0.0508  &[179.7-344.8] & 0.0219 &[123.7-199.3] &0.01241 \\
20   &  [105.3-170.6] & 0.0693  &[89.8-172.6]  & 0.0221 &[61.85-99.6]  &0.01297 \\
30   &  [70.2-113.7]  & 0.0664  &[59.8-115.0]  & 0.0180 &[41.2-66.4]   &0.01111 \\
40   &  [52.65-85.3]  & 0.0498  &[44.9-86.31]  & 0.0124 &[30.9-49.8]   &0.00808 \\
50   &  [42.1-68.2]   & 0.0336  &[35.9-69.0]   & 0.0088 &[24.74-39.8]  &0.00409 \\
60   &  [35.1-56.8]   & 0.0224  &[29.9-57.5]   & 0.0046 &[20.6-33.2]   &0.00201 \\
70   &  [30.1-48.7]   & 0.0146  &[25.6-49.3]   & 0.0031 &[17.67-28.4]  &0.00126 \\
80   &  [26.3-42.6]   & 0.0101  &[22.4-43.1]   & 0.0021 &[15.4-24.9]   &0.00043 \\
90   &  [23.4-37.9]   & 0.0071  &[19.9-38.3]   & 0.0012 &[13.7-22.13]  &0.00021 \\
100  &  [21.06-34.1]  & 0.0047  &[17.9-34.5]   & 0.0008 &[12.37-19.9]  &0.00008 \\
\hline
\end{tabular}
 \caption{The mismatch of numerical waveform SXS:BBH:1410 versus the hybrid of numerical with spin-Taylor T4 approximant. The first column shows the total mass of the binary. We construct three hybrids with hybrid intervals starting 10, 20 and 40 cycles before merger. The two columns for each case show frequency in Hz within the hybrid interval and the weighted mismatch as computed in Eq.~(\ref{eq:WM}). Clearly mismatches are better when one uses  longer numerical waveforms.}
\label{tab:IntervalMismatch}
\end{table*}

\section{Discussion}
\label{sec:discussion}
Hybrid NR waveforms have two potentially direct  applications to GW observations, particularly as the sensitivity of
GW detectors improves at low frequency.  First and foremost, a sufficiently
dense and long family can be directly applied as search templates \cite{2014PhRvD..89d2002K} with a mismatch
target of $0.03$.   For high-mass binaries, hybridization is critical to extend short simulations into a sufficiently
dense and reliable bank. Second, families of NR simulations which reproduce existing candidate events, including potentially directly targeted simulations, can be directly compared to the data,
producing likelihoods for each simulation and mass, along with best-fit GW signals and residuals.    By stitching these
likelihoods together, one can  directly infer the source responsible for the candidate event.
However, both of these analyses are systematically biased by NR simulation's finite durations, when their relevant modes start above the lowest observationally accessible GW frequency.  Hybridization is critical to reduce these effects and enable detection and
parameter inference with NR.

%
The accuracy thresholds for these two applications can be more concretely understood using the conventional mismatch
threshold  required for detection ($0.03$) and to avoid systematic bias in parameter inference [$1/\rho^2$, for $\rho$ the source signal-to-noise ratio (SNR)].  For sources with high red-shifted total mass $M_z=(1+z)M > 100 M_\odot$,  the NR signal alone suffices and
hybridization has relatively little impact: for the strongest mode, mismatches are well below $10^{-5}$  independent of
hybridization, suggesting reliable inference for signals up to $\rho\simeq 300$.  
 For comparison, in a Euclidean cosmology we would need roughly
50 years at a detection rate of 1000/yr to find a source of that magnitude.  Equivalently, for sources with this high
red-shifted mass, NR alone will be more than adequate enough up to the Voyager era.
Conversely, for sources with very low red-shifted mass, hybridization is dominated by inspiral, and the mismatch reflects
systematic differences between GR (as calculated with NR simulations) and our early-time approximations.    In the case
described in this work, we emphasized a PN-based early-time approximation with substantial systematic errors.  
In between these two limits, hybridization generally occurs inside the detector's sensitive band.  Because of the
early-time approximations we employed,  the mismatch generally decreases almost monotonically as source mass increases
and as thus the analyzed signal contains less of the early-time model.   
As a result, for a very loose mismatch threshold of  $10^{-3}$ for parameter inference and using the $(\ell=2, m=2)$ mode
mismatches as key diagnostics, our results suggest even short hybrids with 20 cycles before merger  are
generally reliable above $50 M_\odot$.   Due to large PN differences with NR, hybridizing earlier would not enable
access to significantly lower masses with  high accuracy but would dramatically increase the accuracy of the hybrid at
high mass and thus the ability to use this approach for high-amplitude signals.
Based on prior work, we anticipate that with a superior early-time model, the mismatch would have a local maximum versus mass related to the characteristic frequency at which hybridization was performed.

In general, we see that higher-order modes show larger mismatches than lower-order modes. As both the PN/EOB and NR
errors for these modes are expected to be larger than for the lower-order modes, this is perhaps not
surprising. Interestingly, as shown in Fig.~\ref{fig:StrainEarlyInspiralSXS1392}, despite the relatively large mismatches in these modes,  including these higher-order modes leads to substantially smaller mismatches. As shown  in Fig~\ref{fig:StrainEarlyInspiralSXS1392}, the mismatch is almost 10 times larger when comparing a hybrid constructed 40 cycles prior to merger that uses only the quadrupolar modes (mismatch against the full numerical waveform with all modes) to a hybrid that uses all modes up to the $\ell=5$ modes. When the hybrid is constructed closer to merger, the mismatch is 3 to 4 times larger if only the quadrupolar modes are used.

\section{Conclusion}

We have introduced and assessed a simple, automated algorithm to hybridize gravitational waves from generic precessing
quasicircular binaries.
In this work, we hybridize in an inertial frame, choosing consistent orientations for the pre-
and post-merger binary  such that a waveform-derived estimate of the orbital angular
momentum  $\mathbf{L}$ is along the $z$-axis.   
This simple procedure avoids the need to carefully understand and reproduce precessional dynamics smoothly through the
hybridization interval, which is needed for proposals which  hybridize in a coprecessing frame
We assessed our approach by comparing long NR simulations to  hybrids of artificially-truncated variants of those same simulations.  
As expected, we found that the choice of early-time waveform has significant impact on the quality of the overall
hybrid.  EOB-based hybrids had better behavior at very low-mass; post-Newtonian hybrids, however, showed increasing
mismatch with NR for very low masses, suggesting systematic relative dephasing in long waveforms. 
For generic quasicircular binaries, we were only able to hybridize with existing PN-based approximations, and as a
result our hybrids performed poorly at very low detector-frame mass ($M_z\lesssim 30$), where inspiral dominates the
signal. 
For the very loose mismatch tolerances needed for searches, our hybrid procedure would be  more than sufficient for all
masses investigated here, implying
NR-based searches are limited solely by simulation density.
Conversely, for the tighter mismatch thresholds needed for parameter inference ($1/\rho^2$, typically $10^{-3}-10^{-4}$ for
contemporary observations),  the precessing NR/PN hybrids demonstrated here are expected to be reliable only for red-shifted masses
$M_z>50 M_\odot$, depending somewhat on mass ratio.
We expect hybrids with improved models will produce better performance at early times and low masses.

Hybrid NR waveforms have been applied directly to analyze GW signals.  Already, by mitigating the errors
introduced by abrupt early-time truncation,  our hybridization method will enable even relatively short
NR simulations to be usefully compared to GW observations.   To enable this method to analyze lower masses, however, we
will need early-time models which are more phase-coherent with numerical relativity.  We will explore the impact of alternative
early-time  models in future work.
That said, particularly at the high red-shifted masses $M_z> 100 M_\odot$ which are most relevant to future high red-shift observations of known binary black hole (BBH) populations, our hybrids will be immediately relevant for data analysis even for
high-amplitude signals relevant to the next generation of detectors.

\begin{acknowledgments}
J.S. acknowledges support from Fulbright. Y.Z.
gratefully acknowledges the National Science Foundation (NSF) for
financial support from Grants
No.\ PHY-1707946,  No.\ PHY-1912632, No.\ AST-1516150, No.\ ACI-1550436, 
No.\ PHY-1726215, No.\ PHY-1607520, and No.\ OAC-1811228. Y.Z. gratefully acknowledges NASA for financial support from Grants No.\ NASA NNH17ZDA001N-TCAN-17-TCAN17-0018 80NSSC18K1488 and No.\ NASA 80NSSC20K0528. R.O.S. and J.L. gratefully acknowledge the National Science Foundation (NSF) for
financial support from Grants No. PHY 1707965, No. PHY-1912632, and No. AST-1909534.
This work used the Extreme
Science and Engineering
Discovery Environment (XSEDE) [allocation TG-PHY060027N], which is
supported by NSF Grant No. ACI-1548562.
Computational resources were also provided by the NewHorizons and BlueSky Clusters at the Rochester Institute of Technology, which were
supported by NSF Grants No.\ PHY-0722703, No.\ DMS-0820923, No.\
AST-1028087, and No.\ PHY-1229173.
\end{acknowledgments}

\bibliographystyle{apsrev4-1}
\bibliography{references}

\begin{thebibliography}{65}%
\makeatletter
\providecommand \@ifxundefined [1]{%
 \@ifx{#1\undefined}
}%
\providecommand \@ifnum [1]{%
 \ifnum #1\expandafter \@firstoftwo
 \else \expandafter \@secondoftwo
 \fi
}%
\providecommand \@ifx [1]{%
 \ifx #1\expandafter \@firstoftwo
 \else \expandafter \@secondoftwo
 \fi
}%
\providecommand \natexlab [1]{#1}%
\providecommand \enquote  [1]{``#1''}%
\providecommand \bibnamefont  [1]{#1}%
\providecommand \bibfnamefont [1]{#1}%
\providecommand \citenamefont [1]{#1}%
\providecommand \href@noop [0]{\@secondoftwo}%
\providecommand \href [0]{\begingroup \@sanitize@url \@href}%
\providecommand \@href[1]{\@@startlink{#1}\@@href}%
\providecommand \@@href[1]{\endgroup#1\@@endlink}%
\providecommand \@sanitize@url [0]{\catcode `\\12\catcode `\$12\catcode
  `\&12\catcode `\#12\catcode `\^12\catcode `\_12\catcode `\%12\relax}%
\providecommand \@@startlink[1]{}%
\providecommand \@@endlink[0]{}%
\providecommand \url  [0]{\begingroup\@sanitize@url \@url }%
\providecommand \@url [1]{\endgroup\@href {#1}{\urlprefix }}%
\providecommand \urlprefix  [0]{URL }%
\providecommand \Eprint [0]{\href }%
\providecommand \doibase [0]{http://dx.doi.org/}%
\providecommand \selectlanguage [0]{\@gobble}%
\providecommand \bibinfo  [0]{\@secondoftwo}%
\providecommand \bibfield  [0]{\@secondoftwo}%
\providecommand \translation [1]{[#1]}%
\providecommand \BibitemOpen [0]{}%
\providecommand \bibitemStop [0]{}%
\providecommand \bibitemNoStop [0]{.\EOS\space}%
\providecommand \EOS [0]{\spacefactor3000\relax}%
\providecommand \BibitemShut  [1]{\csname bibitem#1\endcsname}%
\let\auto@bib@innerbib\@empty
\bibitem [{\citenamefont {Abbott}\ \emph
  {et~al.}(2016{\natexlab{a}})\citenamefont {Abbott} \emph
  {et~al.}}]{Abbott:2016blz}%
  \BibitemOpen
  \bibfield  {author} {\bibinfo {author} {\bibfnamefont {B.~P.}\ \bibnamefont
  {Abbott}} \emph {et~al.} (\bibinfo {collaboration} {LIGO Scientific,
  Virgo}),\ }\href {\doibase 10.1103/PhysRevLett.116.061102} {\bibfield
  {journal} {\bibinfo  {journal} {Phys. Rev. Lett.}\ }\textbf {\bibinfo
  {volume} {116}},\ \bibinfo {pages} {061102} (\bibinfo {year}
  {2016}{\natexlab{a}})},\ \Eprint {http://arxiv.org/abs/1602.03837}
  {arXiv:1602.03837 [gr-qc]} \BibitemShut {NoStop}%
\bibitem [{\citenamefont {Dwyer}(2015)}]{Dwyer:2015fua}%
  \BibitemOpen
  \bibfield  {author} {\bibinfo {author} {\bibfnamefont {S.}~\bibnamefont
  {Dwyer}} (\bibinfo {collaboration} {LIGO Scientific}),\ }\bibfield
  {booktitle} {\emph {\bibinfo {booktitle} {{Proceedings, 10th International
  LISA Symposium: Gainesville, Florida, USA, May 18-23, 2014}}},\ }\href
  {\doibase 10.1088/1742-6596/610/1/012013} {\bibfield  {journal} {\bibinfo
  {journal} {J. Phys. Conf. Ser.}\ }\textbf {\bibinfo {volume} {610}},\
  \bibinfo {pages} {012013} (\bibinfo {year} {2015})}\BibitemShut {NoStop}%
\bibitem [{\citenamefont {Heitmann}(2018)}]{Heitmann:2018how}%
  \BibitemOpen
  \bibfield  {author} {\bibinfo {author} {\bibfnamefont {H.}~\bibnamefont
  {Heitmann}} (\bibinfo {collaboration} {VIRGO}),\ }\bibfield  {booktitle}
  {\emph {\bibinfo {booktitle} {{Proceedings, SPIE Astronomical Telescopes +
  Instrumentation 2018: Modeling, Systems Engineering, and Project Management
  for Astronomy VIII: Austin, USA, June 10-15, 2018}}},\ }\href {\doibase
  10.1117/12.2312572} {\bibfield  {journal} {\bibinfo  {journal} {Proc. SPIE
  Int. Soc. Opt. Eng.}\ }\textbf {\bibinfo {volume} {10700}},\ \bibinfo {pages}
  {1070017} (\bibinfo {year} {2018})}\BibitemShut {NoStop}%
\bibitem [{\citenamefont {Akutsu}\ \emph {et~al.}(2019)\citenamefont {Akutsu}
  \emph {et~al.}}]{Akutsu:2018axf}%
  \BibitemOpen
  \bibfield  {author} {\bibinfo {author} {\bibfnamefont {T.}~\bibnamefont
  {Akutsu}} \emph {et~al.} (\bibinfo {collaboration} {KAGRA}),\ }\href
  {\doibase 10.1038/s41550-018-0658-y} {\bibfield  {journal} {\bibinfo
  {journal} {Nat. Astron.}\ }\textbf {\bibinfo {volume} {3}},\ \bibinfo {pages}
  {35} (\bibinfo {year} {2019})},\ \Eprint {http://arxiv.org/abs/1811.08079}
  {arXiv:1811.08079 [gr-qc]} \BibitemShut {NoStop}%
\bibitem [{\citenamefont {Sesana}\ \emph {et~al.}(2014)\citenamefont {Sesana}
  \emph {et~al.}}]{Sesana:2014usa}%
  \BibitemOpen
  \bibfield  {author} {\bibinfo {author} {\bibfnamefont {A.}~\bibnamefont
  {Sesana}} \emph {et~al.},\ }\href {\doibase 10.1007/s10714-014-1793-0}
  {\bibfield  {journal} {\bibinfo  {journal} {Gen. Rel. Grav.}\ }\textbf
  {\bibinfo {volume} {46}},\ \bibinfo {pages} {1793} (\bibinfo {year}
  {2014})}\BibitemShut {NoStop}%
\bibitem [{\citenamefont {McNamara}\ \emph {et~al.}(2013)\citenamefont
  {McNamara} \emph {et~al.}}]{McNamara:2013bma}%
  \BibitemOpen
  \bibfield  {author} {\bibinfo {author} {\bibfnamefont {P.}~\bibnamefont
  {McNamara}} \emph {et~al.},\ }\bibfield  {booktitle} {\emph {\bibinfo
  {booktitle} {{Proceedings, 9th International LISA Symposium (LISA 2012):
  Paris, France, May 21-25, 2012}}},\ }\href@noop {} {\bibfield  {journal}
  {\bibinfo  {journal} {ASP Conf. Ser.}\ }\textbf {\bibinfo {volume} {467}},\
  \bibinfo {pages} {5} (\bibinfo {year} {2013})}\BibitemShut {NoStop}%
\bibitem [{\citenamefont {Abbott}\ \emph
  {et~al.}(2019{\natexlab{a}})\citenamefont {Abbott} \emph
  {et~al.}}]{LIGOScientific:2018mvr}%
  \BibitemOpen
  \bibfield  {author} {\bibinfo {author} {\bibfnamefont {B.~P.}\ \bibnamefont
  {Abbott}} \emph {et~al.} (\bibinfo {collaboration} {LIGO Scientific,
  Virgo}),\ }\href {\doibase 10.1103/PhysRevX.9.031040} {\bibfield  {journal}
  {\bibinfo  {journal} {Phys. Rev.}\ }\textbf {\bibinfo {volume} {X9}},\
  \bibinfo {pages} {031040} (\bibinfo {year} {2019}{\natexlab{a}})},\ \Eprint
  {http://arxiv.org/abs/1811.12907} {arXiv:1811.12907 [astro-ph.HE]}
  \BibitemShut {NoStop}%
\bibitem [{\citenamefont {Abbott}\ \emph
  {et~al.}(2016{\natexlab{b}})\citenamefont {Abbott} \emph
  {et~al.}}]{Abbott:2016nhf}%
  \BibitemOpen
  \bibfield  {author} {\bibinfo {author} {\bibfnamefont {B.~P.}\ \bibnamefont
  {Abbott}} \emph {et~al.} (\bibinfo {collaboration} {LIGO Scientific,
  Virgo}),\ }\href {\doibase 10.3847/2041-8205/833/1/L1} {\bibfield  {journal}
  {\bibinfo  {journal} {Astrophys. J.}\ }\textbf {\bibinfo {volume} {833}},\
  \bibinfo {pages} {L1} (\bibinfo {year} {2016}{\natexlab{b}})},\ \Eprint
  {http://arxiv.org/abs/1602.03842} {arXiv:1602.03842 [astro-ph.HE]}
  \BibitemShut {NoStop}%
\bibitem [{\citenamefont {Abbott}\ \emph
  {et~al.}(2019{\natexlab{b}})\citenamefont {Abbott} \emph
  {et~al.}}]{LIGOScientific:2018jsj}%
  \BibitemOpen
  \bibfield  {author} {\bibinfo {author} {\bibfnamefont {B.~P.}\ \bibnamefont
  {Abbott}} \emph {et~al.} (\bibinfo {collaboration} {LIGO Scientific,
  Virgo}),\ }\href {\doibase 10.3847/2041-8213/ab3800} {\bibfield  {journal}
  {\bibinfo  {journal} {Astrophys. J.}\ }\textbf {\bibinfo {volume} {882}},\
  \bibinfo {pages} {L24} (\bibinfo {year} {2019}{\natexlab{b}})},\ \Eprint
  {http://arxiv.org/abs/1811.12940} {arXiv:1811.12940 [astro-ph.HE]}
  \BibitemShut {NoStop}%
\bibitem [{\citenamefont {Gerosa}\ \emph {et~al.}(2019)\citenamefont {Gerosa},
  \citenamefont {Ma}, \citenamefont {Wong}, \citenamefont {Berti},
  \citenamefont {O'Shaughnessy}, \citenamefont {Chen},\ and\ \citenamefont
  {Belczynski}}]{Gerosa:2019dbe}%
  \BibitemOpen
  \bibfield  {author} {\bibinfo {author} {\bibfnamefont {D.}~\bibnamefont
  {Gerosa}}, \bibinfo {author} {\bibfnamefont {S.}~\bibnamefont {Ma}}, \bibinfo
  {author} {\bibfnamefont {K.~W.~K.}\ \bibnamefont {Wong}}, \bibinfo {author}
  {\bibfnamefont {E.}~\bibnamefont {Berti}}, \bibinfo {author} {\bibfnamefont
  {R.}~\bibnamefont {O'Shaughnessy}}, \bibinfo {author} {\bibfnamefont
  {Y.}~\bibnamefont {Chen}}, \ and\ \bibinfo {author} {\bibfnamefont
  {K.}~\bibnamefont {Belczynski}},\ }\href {\doibase
  10.1103/PhysRevD.99.103004} {\bibfield  {journal} {\bibinfo  {journal} {Phys.
  Rev.}\ }\textbf {\bibinfo {volume} {D99}},\ \bibinfo {pages} {103004}
  (\bibinfo {year} {2019})},\ \Eprint {http://arxiv.org/abs/1902.00021}
  {arXiv:1902.00021 [astro-ph.HE]} \BibitemShut {NoStop}%
\bibitem [{\citenamefont {Cutler}\ \emph {et~al.}(1993)\citenamefont {Cutler}
  \emph {et~al.}}]{Cutler:1992tc}%
  \BibitemOpen
  \bibfield  {author} {\bibinfo {author} {\bibfnamefont {C.}~\bibnamefont
  {Cutler}} \emph {et~al.},\ }\href {\doibase 10.1103/PhysRevLett.70.2984}
  {\bibfield  {journal} {\bibinfo  {journal} {Phys. Rev. Lett.}\ }\textbf
  {\bibinfo {volume} {70}},\ \bibinfo {pages} {2984} (\bibinfo {year}
  {1993})},\ \Eprint {http://arxiv.org/abs/astro-ph/9208005}
  {arXiv:astro-ph/9208005 [astro-ph]} \BibitemShut {NoStop}%
\bibitem [{\citenamefont {Blanchet}(2014)}]{Blanchet:2013haa}%
  \BibitemOpen
  \bibfield  {author} {\bibinfo {author} {\bibfnamefont {L.}~\bibnamefont
  {Blanchet}},\ }\href {\doibase 10.12942/lrr-2014-2} {\bibfield  {journal}
  {\bibinfo  {journal} {Living Rev. Rel.}\ }\textbf {\bibinfo {volume} {17}},\
  \bibinfo {pages} {2} (\bibinfo {year} {2014})},\ \Eprint
  {http://arxiv.org/abs/1310.1528} {arXiv:1310.1528 [gr-qc]} \BibitemShut
  {NoStop}%
\bibitem [{\citenamefont {Centrella}\ \emph {et~al.}(2010)\citenamefont
  {Centrella}, \citenamefont {Baker}, \citenamefont {Kelly},\ and\
  \citenamefont {van Meter}}]{Centrella:2010mx}%
  \BibitemOpen
  \bibfield  {author} {\bibinfo {author} {\bibfnamefont {J.}~\bibnamefont
  {Centrella}}, \bibinfo {author} {\bibfnamefont {J.~G.}\ \bibnamefont
  {Baker}}, \bibinfo {author} {\bibfnamefont {B.~J.}\ \bibnamefont {Kelly}}, \
  and\ \bibinfo {author} {\bibfnamefont {J.~R.}\ \bibnamefont {van Meter}},\
  }\href {\doibase 10.1103/RevModPhys.82.3069} {\bibfield  {journal} {\bibinfo
  {journal} {Rev. Mod. Phys.}\ }\textbf {\bibinfo {volume} {82}},\ \bibinfo
  {pages} {3069} (\bibinfo {year} {2010})},\ \Eprint
  {http://arxiv.org/abs/1010.5260} {arXiv:1010.5260 [gr-qc]} \BibitemShut
  {NoStop}%
\bibitem [{\citenamefont {Sturani}\ \emph {et~al.}(2010)\citenamefont
  {Sturani}, \citenamefont {Fischetti}, \citenamefont {Cadonati}, \citenamefont
  {Guidi}, \citenamefont {Healy}, \citenamefont {Shoemaker},\ and\
  \citenamefont {Vicere}}]{Sturani:2010yv}%
  \BibitemOpen
  \bibfield  {author} {\bibinfo {author} {\bibfnamefont {R.}~\bibnamefont
  {Sturani}}, \bibinfo {author} {\bibfnamefont {S.}~\bibnamefont {Fischetti}},
  \bibinfo {author} {\bibfnamefont {L.}~\bibnamefont {Cadonati}}, \bibinfo
  {author} {\bibfnamefont {G.~M.}\ \bibnamefont {Guidi}}, \bibinfo {author}
  {\bibfnamefont {J.}~\bibnamefont {Healy}}, \bibinfo {author} {\bibfnamefont
  {D.}~\bibnamefont {Shoemaker}}, \ and\ \bibinfo {author} {\bibfnamefont
  {A.}~\bibnamefont {Vicere}},\ }\bibfield  {booktitle} {\emph {\bibinfo
  {booktitle} {{Proceedings, 14th Workshop on Gravitational wave data analysis
  (GWDAW-14): Rome, Italy, January 26-29, 2010}}},\ }\href {\doibase
  10.1088/1742-6596/243/1/012007} {\bibfield  {journal} {\bibinfo  {journal}
  {J. Phys. Conf. Ser.}\ }\textbf {\bibinfo {volume} {243}},\ \bibinfo {pages}
  {012007} (\bibinfo {year} {2010})},\ \Eprint {http://arxiv.org/abs/1005.0551}
  {arXiv:1005.0551 [gr-qc]} \BibitemShut {NoStop}%
\bibitem [{\citenamefont {Khan}\ \emph {et~al.}(2019)\citenamefont {Khan},
  \citenamefont {Chatziioannou}, \citenamefont {Hannam},\ and\ \citenamefont
  {Ohme}}]{Khan:2018fmp}%
  \BibitemOpen
  \bibfield  {author} {\bibinfo {author} {\bibfnamefont {S.}~\bibnamefont
  {Khan}}, \bibinfo {author} {\bibfnamefont {K.}~\bibnamefont {Chatziioannou}},
  \bibinfo {author} {\bibfnamefont {M.}~\bibnamefont {Hannam}}, \ and\ \bibinfo
  {author} {\bibfnamefont {F.}~\bibnamefont {Ohme}},\ }\href {\doibase
  10.1103/PhysRevD.100.024059} {\bibfield  {journal} {\bibinfo  {journal}
  {Phys. Rev.}\ }\textbf {\bibinfo {volume} {D100}},\ \bibinfo {pages} {024059}
  (\bibinfo {year} {2019})},\ \Eprint {http://arxiv.org/abs/1809.10113}
  {arXiv:1809.10113 [gr-qc]} \BibitemShut {NoStop}%
\bibitem [{\citenamefont {Varma}\ \emph {et~al.}(2019)\citenamefont {Varma},
  \citenamefont {Field}, \citenamefont {Scheel}, \citenamefont {Blackman},
  \citenamefont {Kidder},\ and\ \citenamefont {Pfeiffer}}]{Varma:2018mmi}%
  \BibitemOpen
  \bibfield  {author} {\bibinfo {author} {\bibfnamefont {V.}~\bibnamefont
  {Varma}}, \bibinfo {author} {\bibfnamefont {S.~E.}\ \bibnamefont {Field}},
  \bibinfo {author} {\bibfnamefont {M.~A.}\ \bibnamefont {Scheel}}, \bibinfo
  {author} {\bibfnamefont {J.}~\bibnamefont {Blackman}}, \bibinfo {author}
  {\bibfnamefont {L.~E.}\ \bibnamefont {Kidder}}, \ and\ \bibinfo {author}
  {\bibfnamefont {H.~P.}\ \bibnamefont {Pfeiffer}},\ }\href {\doibase
  10.1103/PhysRevD.99.064045} {\bibfield  {journal} {\bibinfo  {journal} {Phys.
  Rev.}\ }\textbf {\bibinfo {volume} {D99}},\ \bibinfo {pages} {064045}
  (\bibinfo {year} {2019})},\ \Eprint {http://arxiv.org/abs/1812.07865}
  {arXiv:1812.07865 [gr-qc]} \BibitemShut {NoStop}%
\bibitem [{\citenamefont {MacDonald}\ \emph {et~al.}(2013)\citenamefont
  {MacDonald}, \citenamefont {Mroue}, \citenamefont {Pfeiffer}, \citenamefont
  {Boyle}, \citenamefont {Kidder}, \citenamefont {Scheel}, \citenamefont
  {Szilagyi},\ and\ \citenamefont {Taylor}}]{MacDonald:2012mp}%
  \BibitemOpen
  \bibfield  {author} {\bibinfo {author} {\bibfnamefont {I.}~\bibnamefont
  {MacDonald}}, \bibinfo {author} {\bibfnamefont {A.~H.}\ \bibnamefont
  {Mroue}}, \bibinfo {author} {\bibfnamefont {H.~P.}\ \bibnamefont {Pfeiffer}},
  \bibinfo {author} {\bibfnamefont {M.}~\bibnamefont {Boyle}}, \bibinfo
  {author} {\bibfnamefont {L.~E.}\ \bibnamefont {Kidder}}, \bibinfo {author}
  {\bibfnamefont {M.~A.}\ \bibnamefont {Scheel}}, \bibinfo {author}
  {\bibfnamefont {B.}~\bibnamefont {Szilagyi}}, \ and\ \bibinfo {author}
  {\bibfnamefont {N.~W.}\ \bibnamefont {Taylor}},\ }\href {\doibase
  10.1103/PhysRevD.87.024009} {\bibfield  {journal} {\bibinfo  {journal} {Phys.
  Rev.}\ }\textbf {\bibinfo {volume} {D87}},\ \bibinfo {pages} {024009}
  (\bibinfo {year} {2013})},\ \Eprint {http://arxiv.org/abs/1210.3007}
  {arXiv:1210.3007 [gr-qc]} \BibitemShut {NoStop}%
\bibitem [{\citenamefont {Varma}\ \emph {et~al.}(2014)\citenamefont {Varma},
  \citenamefont {Ajith}, \citenamefont {Husa}, \citenamefont {Bustillo},
  \citenamefont {Hannam},\ and\ \citenamefont {P\"urrer}}]{PhysRevD.90.124004}%
  \BibitemOpen
  \bibfield  {author} {\bibinfo {author} {\bibfnamefont {V.}~\bibnamefont
  {Varma}}, \bibinfo {author} {\bibfnamefont {P.}~\bibnamefont {Ajith}},
  \bibinfo {author} {\bibfnamefont {S.}~\bibnamefont {Husa}}, \bibinfo {author}
  {\bibfnamefont {J.~C.}\ \bibnamefont {Bustillo}}, \bibinfo {author}
  {\bibfnamefont {M.}~\bibnamefont {Hannam}}, \ and\ \bibinfo {author}
  {\bibfnamefont {M.}~\bibnamefont {P\"urrer}},\ }\href {\doibase
  10.1103/PhysRevD.90.124004} {\bibfield  {journal} {\bibinfo  {journal} {Phys.
  Rev. D}\ }\textbf {\bibinfo {volume} {90}},\ \bibinfo {pages} {124004}
  (\bibinfo {year} {2014})}\BibitemShut {NoStop}%
\bibitem [{\citenamefont {Varma}\ and\ \citenamefont
  {Ajith}(2017)}]{Varma:2016dnf}%
  \BibitemOpen
  \bibfield  {author} {\bibinfo {author} {\bibfnamefont {V.}~\bibnamefont
  {Varma}}\ and\ \bibinfo {author} {\bibfnamefont {P.}~\bibnamefont {Ajith}},\
  }\href {\doibase 10.1103/PhysRevD.96.124024} {\bibfield  {journal} {\bibinfo
  {journal} {Phys. Rev.}\ }\textbf {\bibinfo {volume} {D96}},\ \bibinfo {pages}
  {124024} (\bibinfo {year} {2017})},\ \Eprint
  {http://arxiv.org/abs/1612.05608} {arXiv:1612.05608 [gr-qc]} \BibitemShut
  {NoStop}%
\bibitem [{\citenamefont {Calder\'on~Bustillo}\ \emph
  {et~al.}(2016)\citenamefont {Calder\'on~Bustillo}, \citenamefont {Husa},
  \citenamefont {Sintes},\ and\ \citenamefont {P\"urrer}}]{PhysRevD.93.084019}%
  \BibitemOpen
  \bibfield  {author} {\bibinfo {author} {\bibfnamefont {J.}~\bibnamefont
  {Calder\'on~Bustillo}}, \bibinfo {author} {\bibfnamefont {S.}~\bibnamefont
  {Husa}}, \bibinfo {author} {\bibfnamefont {A.~M.}\ \bibnamefont {Sintes}}, \
  and\ \bibinfo {author} {\bibfnamefont {M.}~\bibnamefont {P\"urrer}},\ }\href
  {\doibase 10.1103/PhysRevD.93.084019} {\bibfield  {journal} {\bibinfo
  {journal} {Phys. Rev. D}\ }\textbf {\bibinfo {volume} {93}},\ \bibinfo
  {pages} {084019} (\bibinfo {year} {2016})}\BibitemShut {NoStop}%
\bibitem [{\citenamefont {MacDonald}\ \emph {et~al.}(2011)\citenamefont
  {MacDonald}, \citenamefont {Nissanke}, \citenamefont {Pfeiffer},\ and\
  \citenamefont {Pfeiffer}}]{MacDonald:2011ne}%
  \BibitemOpen
  \bibfield  {author} {\bibinfo {author} {\bibfnamefont {I.}~\bibnamefont
  {MacDonald}}, \bibinfo {author} {\bibfnamefont {S.}~\bibnamefont {Nissanke}},
  \bibinfo {author} {\bibfnamefont {H.~P.}\ \bibnamefont {Pfeiffer}}, \ and\
  \bibinfo {author} {\bibfnamefont {H.~P.}\ \bibnamefont {Pfeiffer}},\
  }\bibfield  {booktitle} {\emph {\bibinfo {booktitle} {{Theory meets data
  analysis at comparable and extreme mass ratios. Proceedings, Conference,
  NRDA/CAPRA 2010, Waterloo, Canada, June 20-26, 2010}}},\ }\href {\doibase
  10.1088/0264-9381/28/13/134002} {\bibfield  {journal} {\bibinfo  {journal}
  {Class. Quant. Grav.}\ }\textbf {\bibinfo {volume} {28}},\ \bibinfo {pages}
  {134002} (\bibinfo {year} {2011})},\ \Eprint {http://arxiv.org/abs/1102.5128}
  {arXiv:1102.5128 [gr-qc]} \BibitemShut {NoStop}%
\bibitem [{\citenamefont {Hannam}\ \emph {et~al.}(2008)\citenamefont {Hannam},
  \citenamefont {Husa}, \citenamefont {Sperhake}, \citenamefont {Bruegmann},\
  and\ \citenamefont {Gonzalez}}]{Hannam:2007ik}%
  \BibitemOpen
  \bibfield  {author} {\bibinfo {author} {\bibfnamefont {M.}~\bibnamefont
  {Hannam}}, \bibinfo {author} {\bibfnamefont {S.}~\bibnamefont {Husa}},
  \bibinfo {author} {\bibfnamefont {U.}~\bibnamefont {Sperhake}}, \bibinfo
  {author} {\bibfnamefont {B.}~\bibnamefont {Bruegmann}}, \ and\ \bibinfo
  {author} {\bibfnamefont {J.~A.}\ \bibnamefont {Gonzalez}},\ }\href {\doibase
  10.1103/PhysRevD.77.044020} {\bibfield  {journal} {\bibinfo  {journal} {Phys.
  Rev.}\ }\textbf {\bibinfo {volume} {D77}},\ \bibinfo {pages} {044020}
  (\bibinfo {year} {2008})},\ \Eprint {http://arxiv.org/abs/0706.1305}
  {arXiv:0706.1305 [gr-qc]} \BibitemShut {NoStop}%
\bibitem [{\citenamefont {Ajith}(2008)}]{Ajith:2007xh}%
  \BibitemOpen
  \bibfield  {author} {\bibinfo {author} {\bibfnamefont {P.}~\bibnamefont
  {Ajith}},\ }\bibfield  {booktitle} {\emph {\bibinfo {booktitle}
  {{Proceedings, 18th International Conference on General Relativity and
  Gravitation (GRG18) and 7th Edoardo Amaldi Conference on Gravitational Waves
  (Amaldi7), Sydney, Australia, July 2007}}},\ }\href {\doibase
  10.1088/0264-9381/25/11/114033} {\bibfield  {journal} {\bibinfo  {journal}
  {Class. Quant. Grav.}\ }\textbf {\bibinfo {volume} {25}},\ \bibinfo {pages}
  {114033} (\bibinfo {year} {2008})},\ \Eprint {http://arxiv.org/abs/0712.0343}
  {arXiv:0712.0343 [gr-qc]} \BibitemShut {NoStop}%
\bibitem [{\citenamefont {Hannam}\ \emph {et~al.}(2010)\citenamefont {Hannam},
  \citenamefont {Husa}, \citenamefont {Ohme},\ and\ \citenamefont
  {Ajith}}]{Hannam:2010ky}%
  \BibitemOpen
  \bibfield  {author} {\bibinfo {author} {\bibfnamefont {M.}~\bibnamefont
  {Hannam}}, \bibinfo {author} {\bibfnamefont {S.}~\bibnamefont {Husa}},
  \bibinfo {author} {\bibfnamefont {F.}~\bibnamefont {Ohme}}, \ and\ \bibinfo
  {author} {\bibfnamefont {P.}~\bibnamefont {Ajith}},\ }\href {\doibase
  10.1103/PhysRevD.82.124052} {\bibfield  {journal} {\bibinfo  {journal} {Phys.
  Rev.}\ }\textbf {\bibinfo {volume} {D82}},\ \bibinfo {pages} {124052}
  (\bibinfo {year} {2010})},\ \Eprint {http://arxiv.org/abs/1008.2961}
  {arXiv:1008.2961 [gr-qc]} \BibitemShut {NoStop}%
\bibitem [{\citenamefont {{The LIGO Scientific Collaboration}}\ \emph
  {et~al.}(2018)\citenamefont {{The LIGO Scientific Collaboration}},
  \citenamefont {{The Virgo Collaboration}}, \citenamefont {{Abbott}},
  \citenamefont {{Abbott}}, \citenamefont {{Abbott}}, \citenamefont
  {{Acernese}}, \citenamefont {{Ackley}}, \citenamefont {{Adams}},
  \citenamefont {{Adams}}, \citenamefont {{Addesso}},\ and\ \citenamefont
  {et~al.}}]{LIGO-O2-Catalog}%
  \BibitemOpen
  \bibfield  {author} {\bibinfo {author} {\bibnamefont {{The LIGO Scientific
  Collaboration}}}, \bibinfo {author} {\bibnamefont {{The Virgo
  Collaboration}}}, \bibinfo {author} {\bibfnamefont {B.~P.}\ \bibnamefont
  {{Abbott}}}, \bibinfo {author} {\bibfnamefont {R.}~\bibnamefont {{Abbott}}},
  \bibinfo {author} {\bibfnamefont {T.~D.}\ \bibnamefont {{Abbott}}}, \bibinfo
  {author} {\bibfnamefont {F.}~\bibnamefont {{Acernese}}}, \bibinfo {author}
  {\bibfnamefont {K.}~\bibnamefont {{Ackley}}}, \bibinfo {author}
  {\bibfnamefont {C.}~\bibnamefont {{Adams}}}, \bibinfo {author} {\bibfnamefont
  {T.}~\bibnamefont {{Adams}}}, \bibinfo {author} {\bibfnamefont
  {P.}~\bibnamefont {{Addesso}}}, \ and\ \bibinfo {author} {\bibnamefont
  {et~al.}},\ }\href {\doibase 10.1103/PhysRevX.9.031040} {\bibfield  {journal}
  {\bibinfo  {journal} {\prx}\ }\textbf {\bibinfo {volume} {9}},\ \bibinfo
  {eid} {031040} (\bibinfo {year} {2018})}\BibitemShut {NoStop}%
\bibitem [{\citenamefont {{Schmidt}}\ \emph {et~al.}(2012)\citenamefont
  {{Schmidt}}, \citenamefont {{Hannam}},\ and\ \citenamefont
  {{Husa}}}]{2012PhRvD..86j4063S}%
  \BibitemOpen
  \bibfield  {author} {\bibinfo {author} {\bibfnamefont {P.}~\bibnamefont
  {{Schmidt}}}, \bibinfo {author} {\bibfnamefont {M.}~\bibnamefont {{Hannam}}},
  \ and\ \bibinfo {author} {\bibfnamefont {S.}~\bibnamefont {{Husa}}},\ }\href
  {\doibase 10.1103/PhysRevD.86.104063} {\bibfield  {journal} {\bibinfo
  {journal} {\prd}\ }\textbf {\bibinfo {volume} {86}},\ \bibinfo {eid} {104063}
  (\bibinfo {year} {2012})},\ \Eprint {http://arxiv.org/abs/1207.3088}
  {arXiv:1207.3088 [gr-qc]} \BibitemShut {NoStop}%
\bibitem [{\citenamefont {{Klein}}\ \emph {et~al.}(2016)\citenamefont
  {{Klein}}, \citenamefont {{Barausse}}, \citenamefont {{Sesana}},
  \citenamefont {{Petiteau}}, \citenamefont {{Berti}}, \citenamefont {{Babak}},
  \citenamefont {{Gair}}, \citenamefont {{Aoudia}}, \citenamefont {{Hinder}},
  \citenamefont {{Ohme}},\ and\ \citenamefont
  {{Wardell}}}]{2016PhRvD..93b4003K}%
  \BibitemOpen
  \bibfield  {author} {\bibinfo {author} {\bibfnamefont {A.}~\bibnamefont
  {{Klein}}}, \bibinfo {author} {\bibfnamefont {E.}~\bibnamefont {{Barausse}}},
  \bibinfo {author} {\bibfnamefont {A.}~\bibnamefont {{Sesana}}}, \bibinfo
  {author} {\bibfnamefont {A.}~\bibnamefont {{Petiteau}}}, \bibinfo {author}
  {\bibfnamefont {E.}~\bibnamefont {{Berti}}}, \bibinfo {author} {\bibfnamefont
  {S.}~\bibnamefont {{Babak}}}, \bibinfo {author} {\bibfnamefont
  {J.}~\bibnamefont {{Gair}}}, \bibinfo {author} {\bibfnamefont
  {S.}~\bibnamefont {{Aoudia}}}, \bibinfo {author} {\bibfnamefont
  {I.}~\bibnamefont {{Hinder}}}, \bibinfo {author} {\bibfnamefont
  {F.}~\bibnamefont {{Ohme}}}, \ and\ \bibinfo {author} {\bibfnamefont
  {B.}~\bibnamefont {{Wardell}}},\ }\href {\doibase 10.1103/PhysRevD.93.024003}
  {\bibfield  {journal} {\bibinfo  {journal} {\prd}\ }\textbf {\bibinfo
  {volume} {93}},\ \bibinfo {eid} {024003} (\bibinfo {year} {2016})},\ \Eprint
  {http://arxiv.org/abs/1511.05581} {arXiv:1511.05581 [gr-qc]} \BibitemShut
  {NoStop}%
\bibitem [{\citenamefont {{Hannam}}\ \emph {et~al.}(2014)\citenamefont
  {{Hannam}}, \citenamefont {{Schmidt}}, \citenamefont {{Boh{\'e}}},
  \citenamefont {{Haegel}}, \citenamefont {{Husa}}, \citenamefont {{Ohme}},
  \citenamefont {{Pratten}},\ and\ \citenamefont
  {{P{\"u}rrer}}}]{2014PhRvL.113o1101H}%
  \BibitemOpen
  \bibfield  {author} {\bibinfo {author} {\bibfnamefont {M.}~\bibnamefont
  {{Hannam}}}, \bibinfo {author} {\bibfnamefont {P.}~\bibnamefont {{Schmidt}}},
  \bibinfo {author} {\bibfnamefont {A.}~\bibnamefont {{Boh{\'e}}}}, \bibinfo
  {author} {\bibfnamefont {L.}~\bibnamefont {{Haegel}}}, \bibinfo {author}
  {\bibfnamefont {S.}~\bibnamefont {{Husa}}}, \bibinfo {author} {\bibfnamefont
  {F.}~\bibnamefont {{Ohme}}}, \bibinfo {author} {\bibfnamefont
  {G.}~\bibnamefont {{Pratten}}}, \ and\ \bibinfo {author} {\bibfnamefont
  {M.}~\bibnamefont {{P{\"u}rrer}}},\ }\href {\doibase
  10.1103/PhysRevLett.113.151101} {\bibfield  {journal} {\bibinfo  {journal}
  {\prl}\ }\textbf {\bibinfo {volume} {113}},\ \bibinfo {eid} {151101}
  (\bibinfo {year} {2014})},\ \Eprint {http://arxiv.org/abs/1308.3271}
  {arXiv:1308.3271 [gr-qc]} \BibitemShut {NoStop}%
\bibitem [{\citenamefont {Harry}\ \emph {et~al.}(2014)\citenamefont {Harry},
  \citenamefont {Nitz}, \citenamefont {Brown}, \citenamefont {Lundgren},
  \citenamefont {Ochsner},\ and\ \citenamefont {Keppel}}]{Harry:2013tca}%
  \BibitemOpen
  \bibfield  {author} {\bibinfo {author} {\bibfnamefont {I.~W.}\ \bibnamefont
  {Harry}}, \bibinfo {author} {\bibfnamefont {A.~H.}\ \bibnamefont {Nitz}},
  \bibinfo {author} {\bibfnamefont {D.~A.}\ \bibnamefont {Brown}}, \bibinfo
  {author} {\bibfnamefont {A.~P.}\ \bibnamefont {Lundgren}}, \bibinfo {author}
  {\bibfnamefont {E.}~\bibnamefont {Ochsner}}, \ and\ \bibinfo {author}
  {\bibfnamefont {D.}~\bibnamefont {Keppel}},\ }\href {\doibase
  10.1103/PhysRevD.89.024010} {\bibfield  {journal} {\bibinfo  {journal} {Phys.
  Rev.}\ }\textbf {\bibinfo {volume} {D89}},\ \bibinfo {pages} {024010}
  (\bibinfo {year} {2014})},\ \Eprint {http://arxiv.org/abs/1307.3562}
  {arXiv:1307.3562 [gr-qc]} \BibitemShut {NoStop}%
\bibitem [{\citenamefont {Chatziioannou}\ \emph {et~al.}(2014)\citenamefont
  {Chatziioannou}, \citenamefont {Cornish}, \citenamefont {Klein},\ and\
  \citenamefont {Yunes}}]{Chatziioannou:2014bma}%
  \BibitemOpen
  \bibfield  {author} {\bibinfo {author} {\bibfnamefont {K.}~\bibnamefont
  {Chatziioannou}}, \bibinfo {author} {\bibfnamefont {N.}~\bibnamefont
  {Cornish}}, \bibinfo {author} {\bibfnamefont {A.}~\bibnamefont {Klein}}, \
  and\ \bibinfo {author} {\bibfnamefont {N.}~\bibnamefont {Yunes}},\ }\href
  {\doibase 10.1103/PhysRevD.89.104023} {\bibfield  {journal} {\bibinfo
  {journal} {Phys. Rev.}\ }\textbf {\bibinfo {volume} {D89}},\ \bibinfo {pages}
  {104023} (\bibinfo {year} {2014})},\ \Eprint {http://arxiv.org/abs/1404.3180}
  {arXiv:1404.3180 [gr-qc]} \BibitemShut {NoStop}%
\bibitem [{\citenamefont {Dal~Canton}\ \emph {et~al.}(2015)\citenamefont
  {Dal~Canton}, \citenamefont {Lundgren},\ and\ \citenamefont
  {Nielsen}}]{Canton:2014uja}%
  \BibitemOpen
  \bibfield  {author} {\bibinfo {author} {\bibfnamefont {T.}~\bibnamefont
  {Dal~Canton}}, \bibinfo {author} {\bibfnamefont {A.~P.}\ \bibnamefont
  {Lundgren}}, \ and\ \bibinfo {author} {\bibfnamefont {A.~B.}\ \bibnamefont
  {Nielsen}},\ }\href {\doibase 10.1103/PhysRevD.91.062010} {\bibfield
  {journal} {\bibinfo  {journal} {Phys. Rev.}\ }\textbf {\bibinfo {volume}
  {D91}},\ \bibinfo {pages} {062010} (\bibinfo {year} {2015})},\ \Eprint
  {http://arxiv.org/abs/1411.6815} {arXiv:1411.6815 [gr-qc]} \BibitemShut
  {NoStop}%
\bibitem [{\citenamefont {Schmidt}\ \emph {et~al.}(2012)\citenamefont
  {Schmidt}, \citenamefont {Hannam},\ and\ \citenamefont
  {Husa}}]{Schmidt:2012rh}%
  \BibitemOpen
  \bibfield  {author} {\bibinfo {author} {\bibfnamefont {P.}~\bibnamefont
  {Schmidt}}, \bibinfo {author} {\bibfnamefont {M.}~\bibnamefont {Hannam}}, \
  and\ \bibinfo {author} {\bibfnamefont {S.}~\bibnamefont {Husa}},\ }\href
  {\doibase 10.1103/PhysRevD.86.104063} {\bibfield  {journal} {\bibinfo
  {journal} {Phys. Rev.}\ }\textbf {\bibinfo {volume} {D86}},\ \bibinfo {pages}
  {104063} (\bibinfo {year} {2012})},\ \Eprint {http://arxiv.org/abs/1207.3088}
  {arXiv:1207.3088 [gr-qc]} \BibitemShut {NoStop}%
\bibitem [{\citenamefont {Mroue}\ \emph
  {et~al.}(2019{\natexlab{a}})\citenamefont {Mroue}, \citenamefont {Boyle},
  \citenamefont {Lovelace}, \citenamefont {Szilagyi}, \citenamefont {Pfeiffer},
  \citenamefont {Zenginoglu}, \citenamefont {Kidder}, \citenamefont {Taylor},
  \citenamefont {Hemberger},\ and\ \citenamefont
  {Scheel}}]{mroue_abdul_2019_3312153}%
  \BibitemOpen
  \bibfield  {author} {\bibinfo {author} {\bibfnamefont {A.}~\bibnamefont
  {Mroue}}, \bibinfo {author} {\bibfnamefont {M.}~\bibnamefont {Boyle}},
  \bibinfo {author} {\bibfnamefont {G.}~\bibnamefont {Lovelace}}, \bibinfo
  {author} {\bibfnamefont {B.}~\bibnamefont {Szilagyi}}, \bibinfo {author}
  {\bibfnamefont {H.}~\bibnamefont {Pfeiffer}}, \bibinfo {author}
  {\bibfnamefont {A.}~\bibnamefont {Zenginoglu}}, \bibinfo {author}
  {\bibfnamefont {L.}~\bibnamefont {Kidder}}, \bibinfo {author} {\bibfnamefont
  {N.}~\bibnamefont {Taylor}}, \bibinfo {author} {\bibfnamefont
  {D.}~\bibnamefont {Hemberger}}, \ and\ \bibinfo {author} {\bibfnamefont
  {M.}~\bibnamefont {Scheel}},\ }\href {\doibase 10.5281/zenodo.3312153}
  {\enquote {\bibinfo {title} {Binary black-hole simulation sxs:bbh:0058},}\ }
  (\bibinfo {year} {2019}{\natexlab{a}})\BibitemShut {NoStop}%
\bibitem [{\citenamefont {Schmidt}\ \emph {et~al.}(2011)\citenamefont
  {Schmidt}, \citenamefont {Hannam}, \citenamefont {Husa},\ and\ \citenamefont
  {Ajith}}]{PhysRevD.84.024046}%
  \BibitemOpen
  \bibfield  {author} {\bibinfo {author} {\bibfnamefont {P.}~\bibnamefont
  {Schmidt}}, \bibinfo {author} {\bibfnamefont {M.}~\bibnamefont {Hannam}},
  \bibinfo {author} {\bibfnamefont {S.}~\bibnamefont {Husa}}, \ and\ \bibinfo
  {author} {\bibfnamefont {P.}~\bibnamefont {Ajith}},\ }\href {\doibase
  10.1103/PhysRevD.84.024046} {\bibfield  {journal} {\bibinfo  {journal} {Phys.
  Rev. D}\ }\textbf {\bibinfo {volume} {84}},\ \bibinfo {pages} {024046}
  (\bibinfo {year} {2011})}\BibitemShut {NoStop}%
\bibitem [{\citenamefont {O'Shaughnessy}\ \emph {et~al.}(2011)\citenamefont
  {O'Shaughnessy}, \citenamefont {Vaishnav}, \citenamefont {Healy},
  \citenamefont {Meeks},\ and\ \citenamefont
  {Shoemaker}}]{OShaughnessy:2011pmr}%
  \BibitemOpen
  \bibfield  {author} {\bibinfo {author} {\bibfnamefont {R.}~\bibnamefont
  {O'Shaughnessy}}, \bibinfo {author} {\bibfnamefont {B.}~\bibnamefont
  {Vaishnav}}, \bibinfo {author} {\bibfnamefont {J.}~\bibnamefont {Healy}},
  \bibinfo {author} {\bibfnamefont {Z.}~\bibnamefont {Meeks}}, \ and\ \bibinfo
  {author} {\bibfnamefont {D.}~\bibnamefont {Shoemaker}},\ }\href {\doibase
  10.1103/PhysRevD.84.124002} {\bibfield  {journal} {\bibinfo  {journal} {Phys.
  Rev.}\ }\textbf {\bibinfo {volume} {D84}},\ \bibinfo {pages} {124002}
  (\bibinfo {year} {2011})},\ \Eprint {http://arxiv.org/abs/1109.5224}
  {arXiv:1109.5224 [gr-qc]} \BibitemShut {NoStop}%
\bibitem [{\citenamefont {Boyle}\ \emph {et~al.}(2011)\citenamefont {Boyle},
  \citenamefont {Owen},\ and\ \citenamefont {Pfeiffer}}]{Boyle:2011gg}%
  \BibitemOpen
  \bibfield  {author} {\bibinfo {author} {\bibfnamefont {M.}~\bibnamefont
  {Boyle}}, \bibinfo {author} {\bibfnamefont {R.}~\bibnamefont {Owen}}, \ and\
  \bibinfo {author} {\bibfnamefont {H.~P.}\ \bibnamefont {Pfeiffer}},\ }\href
  {\doibase 10.1103/PhysRevD.84.124011} {\bibfield  {journal} {\bibinfo
  {journal} {Phys. Rev.}\ }\textbf {\bibinfo {volume} {D84}},\ \bibinfo {pages}
  {124011} (\bibinfo {year} {2011})},\ \Eprint {http://arxiv.org/abs/1110.2965}
  {arXiv:1110.2965 [gr-qc]} \BibitemShut {NoStop}%
\bibitem [{\citenamefont {Jones}\ \emph {et~al.}(01  )\citenamefont {Jones},
  \citenamefont {Oliphant}, \citenamefont {Peterson} \emph
  {et~al.}}]{Scipy:Py}%
  \BibitemOpen
  \bibfield  {author} {\bibinfo {author} {\bibfnamefont {E.}~\bibnamefont
  {Jones}}, \bibinfo {author} {\bibfnamefont {T.}~\bibnamefont {Oliphant}},
  \bibinfo {author} {\bibfnamefont {P.}~\bibnamefont {Peterson}},  \emph
  {et~al.},\ }\href {http://www.scipy.org/} {\enquote {\bibinfo {title}
  {{SciPy}: Open source scientific tools for {Python}},}\ } (\bibinfo {year}
  {2001--}),\ \bibinfo {note} {[Online; accessed <today>]}\BibitemShut
  {NoStop}%
\bibitem [{\citenamefont {McKechan}\ \emph {et~al.}(2010)\citenamefont
  {McKechan}, \citenamefont {Robinson},\ and\ \citenamefont
  {Sathyaprakash}}]{McKechan:2010kp}%
  \BibitemOpen
  \bibfield  {author} {\bibinfo {author} {\bibfnamefont {D.~J.~A.}\
  \bibnamefont {McKechan}}, \bibinfo {author} {\bibfnamefont {C.}~\bibnamefont
  {Robinson}}, \ and\ \bibinfo {author} {\bibfnamefont {B.~S.}\ \bibnamefont
  {Sathyaprakash}},\ }\bibfield  {booktitle} {\emph {\bibinfo {booktitle}
  {{Gravitational waves. Proceedings, 8th Edoardo Amaldi Conference, Amaldi 8,
  New York, USA, June 22-26, 2009}}},\ }\href {\doibase
  10.1088/0264-9381/27/8/084020} {\bibfield  {journal} {\bibinfo  {journal}
  {Class. Quant. Grav.}\ }\textbf {\bibinfo {volume} {27}},\ \bibinfo {pages}
  {084020} (\bibinfo {year} {2010})},\ \Eprint {http://arxiv.org/abs/1003.2939}
  {arXiv:1003.2939 [gr-qc]} \BibitemShut {NoStop}%
\bibitem [{SXS()}]{SXS:catalog}%
  \BibitemOpen
  \href@noop {} {}\bibinfo {howpublished}
  {\url{http://www.black-holes.org/waveforms}}\BibitemShut {NoStop}%
\bibitem [{RIT()}]{RIT:catalog}%
  \BibitemOpen
  \href@noop {} {}\bibinfo {howpublished}
  {\url{http://ccrgpages.rit.edu/~RITCatalog/}}\BibitemShut {NoStop}%
\bibitem [{\citenamefont {Mroue}\ \emph {et~al.}(2013)\citenamefont {Mroue}
  \emph {et~al.}}]{Mroue:2013xna}%
  \BibitemOpen
  \bibfield  {author} {\bibinfo {author} {\bibfnamefont {A.~H.}\ \bibnamefont
  {Mroue}} \emph {et~al.},\ }\href {\doibase 10.1103/PhysRevLett.111.241104}
  {\bibfield  {journal} {\bibinfo  {journal} {Phys. Rev. Lett.}\ }\textbf
  {\bibinfo {volume} {111}},\ \bibinfo {pages} {241104} (\bibinfo {year}
  {2013})},\ \Eprint {http://arxiv.org/abs/1304.6077} {arXiv:1304.6077 [gr-qc]}
  \BibitemShut {NoStop}%
\bibitem [{\citenamefont {Healy}\ \emph {et~al.}(2019)\citenamefont {Healy},
  \citenamefont {Lousto}, \citenamefont {Lange}, \citenamefont {O'Shaughnessy},
  \citenamefont {Zlochower},\ and\ \citenamefont {Campanelli}}]{Healy:2019jyf}%
  \BibitemOpen
  \bibfield  {author} {\bibinfo {author} {\bibfnamefont {J.}~\bibnamefont
  {Healy}}, \bibinfo {author} {\bibfnamefont {C.~O.}\ \bibnamefont {Lousto}},
  \bibinfo {author} {\bibfnamefont {J.}~\bibnamefont {Lange}}, \bibinfo
  {author} {\bibfnamefont {R.}~\bibnamefont {O'Shaughnessy}}, \bibinfo {author}
  {\bibfnamefont {Y.}~\bibnamefont {Zlochower}}, \ and\ \bibinfo {author}
  {\bibfnamefont {M.}~\bibnamefont {Campanelli}},\ }\href {\doibase
  10.1103/PhysRevD.100.024021} {\bibfield  {journal} {\bibinfo  {journal}
  {Phys. Rev.}\ }\textbf {\bibinfo {volume} {D100}},\ \bibinfo {pages} {024021}
  (\bibinfo {year} {2019})},\ \Eprint {http://arxiv.org/abs/1901.02553}
  {arXiv:1901.02553 [gr-qc]} \BibitemShut {NoStop}%
\bibitem [{\citenamefont {Healy}\ \emph {et~al.}(2017)\citenamefont {Healy},
  \citenamefont {Lousto}, \citenamefont {Zlochower},\ and\ \citenamefont
  {Campanelli}}]{Healy:2017psd}%
  \BibitemOpen
  \bibfield  {author} {\bibinfo {author} {\bibfnamefont {J.}~\bibnamefont
  {Healy}}, \bibinfo {author} {\bibfnamefont {C.~O.}\ \bibnamefont {Lousto}},
  \bibinfo {author} {\bibfnamefont {Y.}~\bibnamefont {Zlochower}}, \ and\
  \bibinfo {author} {\bibfnamefont {M.}~\bibnamefont {Campanelli}},\ }\href
  {\doibase 10.1088/1361-6382/aa91b1} {\bibfield  {journal} {\bibinfo
  {journal} {Class. Quant. Grav.}\ }\textbf {\bibinfo {volume} {34}},\ \bibinfo
  {pages} {224001} (\bibinfo {year} {2017})},\ \Eprint
  {http://arxiv.org/abs/1703.03423} {arXiv:1703.03423 [gr-qc]} \BibitemShut
  {NoStop}%
\bibitem [{\citenamefont {Ajith}(2011)}]{Ajith:2011ec}%
  \BibitemOpen
  \bibfield  {author} {\bibinfo {author} {\bibfnamefont {P.}~\bibnamefont
  {Ajith}},\ }\href {\doibase 10.1103/PhysRevD.84.084037} {\bibfield  {journal}
  {\bibinfo  {journal} {Phys. Rev.}\ }\textbf {\bibinfo {volume} {D84}},\
  \bibinfo {pages} {084037} (\bibinfo {year} {2011})},\ \Eprint
  {http://arxiv.org/abs/1107.1267} {arXiv:1107.1267 [gr-qc]} \BibitemShut
  {NoStop}%
\bibitem [{\citenamefont {Boyle}\ \emph {et~al.}(2014)\citenamefont {Boyle},
  \citenamefont {Kidder}, \citenamefont {Ossokine},\ and\ \citenamefont
  {Pfeiffer}}]{Boyle:2014ioa}%
  \BibitemOpen
  \bibfield  {author} {\bibinfo {author} {\bibfnamefont {M.}~\bibnamefont
  {Boyle}}, \bibinfo {author} {\bibfnamefont {L.~E.}\ \bibnamefont {Kidder}},
  \bibinfo {author} {\bibfnamefont {S.}~\bibnamefont {Ossokine}}, \ and\
  \bibinfo {author} {\bibfnamefont {H.~P.}\ \bibnamefont {Pfeiffer}},\
  }\href@noop {} {\  (\bibinfo {year} {2014})},\ \Eprint
  {http://arxiv.org/abs/1409.4431} {arXiv:1409.4431 [gr-qc]} \BibitemShut
  {NoStop}%
\bibitem [{\citenamefont {Marsat}\ \emph {et~al.}(2014)\citenamefont {Marsat},
  \citenamefont {Bohé}, \citenamefont {Blanchet},\ and\ \citenamefont
  {Buonanno}}]{Marsat:2013caa}%
  \BibitemOpen
  \bibfield  {author} {\bibinfo {author} {\bibfnamefont {S.}~\bibnamefont
  {Marsat}}, \bibinfo {author} {\bibfnamefont {A.}~\bibnamefont {Bohé}},
  \bibinfo {author} {\bibfnamefont {L.}~\bibnamefont {Blanchet}}, \ and\
  \bibinfo {author} {\bibfnamefont {A.}~\bibnamefont {Buonanno}},\ }\href
  {\doibase 10.1088/0264-9381/31/2/025023} {\bibfield  {journal} {\bibinfo
  {journal} {Class. Quant. Grav.}\ }\textbf {\bibinfo {volume} {31}},\ \bibinfo
  {pages} {025023} (\bibinfo {year} {2014})},\ \Eprint
  {http://arxiv.org/abs/1307.6793} {arXiv:1307.6793 [gr-qc]} \BibitemShut
  {NoStop}%
\bibitem [{\citenamefont {Bohé}\ \emph {et~al.}(2015)\citenamefont {Bohé},
  \citenamefont {Faye}, \citenamefont {Marsat},\ and\ \citenamefont
  {Porter}}]{Bohe:2015ana}%
  \BibitemOpen
  \bibfield  {author} {\bibinfo {author} {\bibfnamefont {A.}~\bibnamefont
  {Bohé}}, \bibinfo {author} {\bibfnamefont {G.}~\bibnamefont {Faye}},
  \bibinfo {author} {\bibfnamefont {S.}~\bibnamefont {Marsat}}, \ and\ \bibinfo
  {author} {\bibfnamefont {E.~K.}\ \bibnamefont {Porter}},\ }\href {\doibase
  10.1088/0264-9381/32/19/195010} {\bibfield  {journal} {\bibinfo  {journal}
  {Class. Quant. Grav.}\ }\textbf {\bibinfo {volume} {32}},\ \bibinfo {pages}
  {195010} (\bibinfo {year} {2015})},\ \Eprint
  {http://arxiv.org/abs/1501.01529} {arXiv:1501.01529 [gr-qc]} \BibitemShut
  {NoStop}%
\bibitem [{\citenamefont {Bohe}\ \emph {et~al.}(2013)\citenamefont {Bohe},
  \citenamefont {Marsat}, \citenamefont {Faye},\ and\ \citenamefont
  {Blanchet}}]{Bohe:2012mr}%
  \BibitemOpen
  \bibfield  {author} {\bibinfo {author} {\bibfnamefont {A.}~\bibnamefont
  {Bohe}}, \bibinfo {author} {\bibfnamefont {S.}~\bibnamefont {Marsat}},
  \bibinfo {author} {\bibfnamefont {G.}~\bibnamefont {Faye}}, \ and\ \bibinfo
  {author} {\bibfnamefont {L.}~\bibnamefont {Blanchet}},\ }\href {\doibase
  10.1088/0264-9381/30/7/075017} {\bibfield  {journal} {\bibinfo  {journal}
  {Class. Quant. Grav.}\ }\textbf {\bibinfo {volume} {30}},\ \bibinfo {pages}
  {075017} (\bibinfo {year} {2013})},\ \Eprint {http://arxiv.org/abs/1212.5520}
  {arXiv:1212.5520 [gr-qc]} \BibitemShut {NoStop}%
\bibitem [{\citenamefont {Marsat}(2015)}]{Marsat:2014xea}%
  \BibitemOpen
  \bibfield  {author} {\bibinfo {author} {\bibfnamefont {S.}~\bibnamefont
  {Marsat}},\ }\href {\doibase 10.1088/0264-9381/32/8/085008} {\bibfield
  {journal} {\bibinfo  {journal} {Class. Quant. Grav.}\ }\textbf {\bibinfo
  {volume} {32}},\ \bibinfo {pages} {085008} (\bibinfo {year} {2015})},\
  \Eprint {http://arxiv.org/abs/1411.4118} {arXiv:1411.4118 [gr-qc]}
  \BibitemShut {NoStop}%
\bibitem [{\citenamefont {Levi}\ and\ \citenamefont
  {Steinhoff}(2015)}]{Levi:2014gsa}%
  \BibitemOpen
  \bibfield  {author} {\bibinfo {author} {\bibfnamefont {M.}~\bibnamefont
  {Levi}}\ and\ \bibinfo {author} {\bibfnamefont {J.}~\bibnamefont
  {Steinhoff}},\ }\href {\doibase 10.1007/JHEP06(2015)059} {\bibfield
  {journal} {\bibinfo  {journal} {JHEP}\ }\textbf {\bibinfo {volume} {06}},\
  \bibinfo {pages} {059} (\bibinfo {year} {2015})},\ \Eprint
  {http://arxiv.org/abs/1410.2601} {arXiv:1410.2601 [gr-qc]} \BibitemShut
  {NoStop}%
\bibitem [{\citenamefont {{LIGO Scientific Collaboration}}(2018)}]{lalsuite}%
  \BibitemOpen
  \bibfield  {author} {\bibinfo {author} {\bibnamefont {{LIGO Scientific
  Collaboration}}},\ }\href {\doibase 10.7935/GT1W-FZ16} {\enquote {\bibinfo
  {title} {{LIGO} {A}lgorithm {L}ibrary - {LALS}uite},}\ }\bibinfo
  {howpublished} {free software (GPL)} (\bibinfo {year} {2018})\BibitemShut
  {NoStop}%
\bibitem [{\citenamefont {Pan}\ \emph {et~al.}(2011)\citenamefont {Pan},
  \citenamefont {Buonanno}, \citenamefont {Boyle}, \citenamefont {Buchman},
  \citenamefont {Kidder}, \citenamefont {Pfeiffer},\ and\ \citenamefont
  {Scheel}}]{Pan:2011gk}%
  \BibitemOpen
  \bibfield  {author} {\bibinfo {author} {\bibfnamefont {Y.}~\bibnamefont
  {Pan}}, \bibinfo {author} {\bibfnamefont {A.}~\bibnamefont {Buonanno}},
  \bibinfo {author} {\bibfnamefont {M.}~\bibnamefont {Boyle}}, \bibinfo
  {author} {\bibfnamefont {L.~T.}\ \bibnamefont {Buchman}}, \bibinfo {author}
  {\bibfnamefont {L.~E.}\ \bibnamefont {Kidder}}, \bibinfo {author}
  {\bibfnamefont {H.~P.}\ \bibnamefont {Pfeiffer}}, \ and\ \bibinfo {author}
  {\bibfnamefont {M.~A.}\ \bibnamefont {Scheel}},\ }\href {\doibase
  10.1103/PhysRevD.84.124052} {\bibfield  {journal} {\bibinfo  {journal} {Phys.
  Rev.}\ }\textbf {\bibinfo {volume} {D84}},\ \bibinfo {pages} {124052}
  (\bibinfo {year} {2011})},\ \Eprint {http://arxiv.org/abs/1106.1021}
  {arXiv:1106.1021 [gr-qc]} \BibitemShut {NoStop}%
\bibitem [{\citenamefont {Pan}\ \emph {et~al.}(2010)\citenamefont {Pan},
  \citenamefont {Buonanno}, \citenamefont {Buchman}, \citenamefont {Chu},
  \citenamefont {Kidder}, \citenamefont {Pfeiffer},\ and\ \citenamefont
  {Scheel}}]{Pan:2009wj}%
  \BibitemOpen
  \bibfield  {author} {\bibinfo {author} {\bibfnamefont {Y.}~\bibnamefont
  {Pan}}, \bibinfo {author} {\bibfnamefont {A.}~\bibnamefont {Buonanno}},
  \bibinfo {author} {\bibfnamefont {L.~T.}\ \bibnamefont {Buchman}}, \bibinfo
  {author} {\bibfnamefont {T.}~\bibnamefont {Chu}}, \bibinfo {author}
  {\bibfnamefont {L.~E.}\ \bibnamefont {Kidder}}, \bibinfo {author}
  {\bibfnamefont {H.~P.}\ \bibnamefont {Pfeiffer}}, \ and\ \bibinfo {author}
  {\bibfnamefont {M.~A.}\ \bibnamefont {Scheel}},\ }\href {\doibase
  10.1103/PhysRevD.81.084041} {\bibfield  {journal} {\bibinfo  {journal} {Phys.
  Rev.}\ }\textbf {\bibinfo {volume} {D81}},\ \bibinfo {pages} {084041}
  (\bibinfo {year} {2010})},\ \Eprint {http://arxiv.org/abs/0912.3466}
  {arXiv:0912.3466 [gr-qc]} \BibitemShut {NoStop}%
\bibitem [{\citenamefont {Taracchini}\ \emph {et~al.}(2012)\citenamefont
  {Taracchini}, \citenamefont {Pan}, \citenamefont {Buonanno}, \citenamefont
  {Barausse}, \citenamefont {Boyle}, \citenamefont {Chu}, \citenamefont
  {Lovelace}, \citenamefont {Pfeiffer},\ and\ \citenamefont
  {Scheel}}]{Taracchini:2012ig}%
  \BibitemOpen
  \bibfield  {author} {\bibinfo {author} {\bibfnamefont {A.}~\bibnamefont
  {Taracchini}}, \bibinfo {author} {\bibfnamefont {Y.}~\bibnamefont {Pan}},
  \bibinfo {author} {\bibfnamefont {A.}~\bibnamefont {Buonanno}}, \bibinfo
  {author} {\bibfnamefont {E.}~\bibnamefont {Barausse}}, \bibinfo {author}
  {\bibfnamefont {M.}~\bibnamefont {Boyle}}, \bibinfo {author} {\bibfnamefont
  {T.}~\bibnamefont {Chu}}, \bibinfo {author} {\bibfnamefont {G.}~\bibnamefont
  {Lovelace}}, \bibinfo {author} {\bibfnamefont {H.~P.}\ \bibnamefont
  {Pfeiffer}}, \ and\ \bibinfo {author} {\bibfnamefont {M.~A.}\ \bibnamefont
  {Scheel}},\ }\href {\doibase 10.1103/PhysRevD.86.024011} {\bibfield
  {journal} {\bibinfo  {journal} {Phys. Rev.}\ }\textbf {\bibinfo {volume}
  {D86}},\ \bibinfo {pages} {024011} (\bibinfo {year} {2012})},\ \Eprint
  {http://arxiv.org/abs/1202.0790} {arXiv:1202.0790 [gr-qc]} \BibitemShut
  {NoStop}%
\bibitem [{\citenamefont {Taracchini}\ \emph {et~al.}(2014)\citenamefont
  {Taracchini} \emph {et~al.}}]{Taracchini:2013rva}%
  \BibitemOpen
  \bibfield  {author} {\bibinfo {author} {\bibfnamefont {A.}~\bibnamefont
  {Taracchini}} \emph {et~al.},\ }\href {\doibase 10.1103/PhysRevD.89.061502}
  {\bibfield  {journal} {\bibinfo  {journal} {Phys. Rev.}\ }\textbf {\bibinfo
  {volume} {D89}},\ \bibinfo {pages} {061502} (\bibinfo {year} {2014})},\
  \Eprint {http://arxiv.org/abs/1311.2544} {arXiv:1311.2544 [gr-qc]}
  \BibitemShut {NoStop}%
\bibitem [{\citenamefont {Pan}\ \emph {et~al.}(2014)\citenamefont {Pan},
  \citenamefont {Buonanno}, \citenamefont {Taracchini}, \citenamefont {Kidder},
  \citenamefont {Mrou\'{e}}, \citenamefont {Pfeiffer}, \citenamefont {Scheel},\
  and\ \citenamefont {Szil\'{a}gyi}}]{Pan:2013rra}%
  \BibitemOpen
  \bibfield  {author} {\bibinfo {author} {\bibfnamefont {Y.}~\bibnamefont
  {Pan}}, \bibinfo {author} {\bibfnamefont {A.}~\bibnamefont {Buonanno}},
  \bibinfo {author} {\bibfnamefont {A.}~\bibnamefont {Taracchini}}, \bibinfo
  {author} {\bibfnamefont {L.~E.}\ \bibnamefont {Kidder}}, \bibinfo {author}
  {\bibfnamefont {A.~H.}\ \bibnamefont {Mrou\'{e}}}, \bibinfo {author}
  {\bibfnamefont {H.~P.}\ \bibnamefont {Pfeiffer}}, \bibinfo {author}
  {\bibfnamefont {M.~A.}\ \bibnamefont {Scheel}}, \ and\ \bibinfo {author}
  {\bibfnamefont {B.}~\bibnamefont {Szil\'{a}gyi}},\ }\href {\doibase
  10.1103/PhysRevD.89.084006} {\bibfield  {journal} {\bibinfo  {journal} {Phys.
  Rev.}\ }\textbf {\bibinfo {volume} {D89}},\ \bibinfo {pages} {084006}
  (\bibinfo {year} {2014})},\ \Eprint {http://arxiv.org/abs/1307.6232}
  {arXiv:1307.6232 [gr-qc]} \BibitemShut {NoStop}%
\bibitem [{\citenamefont {Cotesta}\ \emph {et~al.}(2018)\citenamefont
  {Cotesta}, \citenamefont {Buonanno}, \citenamefont {Boh\'e}, \citenamefont
  {Taracchini}, \citenamefont {Hinder},\ and\ \citenamefont
  {Ossokine}}]{PhysRevD.98.084028}%
  \BibitemOpen
  \bibfield  {author} {\bibinfo {author} {\bibfnamefont {R.}~\bibnamefont
  {Cotesta}}, \bibinfo {author} {\bibfnamefont {A.}~\bibnamefont {Buonanno}},
  \bibinfo {author} {\bibfnamefont {A.}~\bibnamefont {Boh\'e}}, \bibinfo
  {author} {\bibfnamefont {A.}~\bibnamefont {Taracchini}}, \bibinfo {author}
  {\bibfnamefont {I.}~\bibnamefont {Hinder}}, \ and\ \bibinfo {author}
  {\bibfnamefont {S.}~\bibnamefont {Ossokine}},\ }\href {\doibase
  10.1103/PhysRevD.98.084028} {\bibfield  {journal} {\bibinfo  {journal} {Phys.
  Rev. D}\ }\textbf {\bibinfo {volume} {98}},\ \bibinfo {pages} {084028}
  (\bibinfo {year} {2018})}\BibitemShut {NoStop}%
\bibitem [{\citenamefont {Mroue}\ \emph
  {et~al.}(2019{\natexlab{b}})\citenamefont {Mroue}, \citenamefont {Boyle},
  \citenamefont {Lovelace}, \citenamefont {Szilagyi}, \citenamefont {Pfeiffer},
  \citenamefont {Zenginoglu}, \citenamefont {Kidder}, \citenamefont {Taylor},
  \citenamefont {Hemberger},\ and\ \citenamefont
  {Scheel}}]{mroue_abdul_2019_3312192}%
  \BibitemOpen
  \bibfield  {author} {\bibinfo {author} {\bibfnamefont {A.}~\bibnamefont
  {Mroue}}, \bibinfo {author} {\bibfnamefont {M.}~\bibnamefont {Boyle}},
  \bibinfo {author} {\bibfnamefont {G.}~\bibnamefont {Lovelace}}, \bibinfo
  {author} {\bibfnamefont {B.}~\bibnamefont {Szilagyi}}, \bibinfo {author}
  {\bibfnamefont {H.}~\bibnamefont {Pfeiffer}}, \bibinfo {author}
  {\bibfnamefont {A.}~\bibnamefont {Zenginoglu}}, \bibinfo {author}
  {\bibfnamefont {L.}~\bibnamefont {Kidder}}, \bibinfo {author} {\bibfnamefont
  {N.}~\bibnamefont {Taylor}}, \bibinfo {author} {\bibfnamefont
  {D.}~\bibnamefont {Hemberger}}, \ and\ \bibinfo {author} {\bibfnamefont
  {M.}~\bibnamefont {Scheel}},\ }\href {\doibase 10.5281/zenodo.3312192}
  {\enquote {\bibinfo {title} {Binary black-hole simulation sxs:bbh:0056},}\ }
  (\bibinfo {year} {2019}{\natexlab{b}})\BibitemShut {NoStop}%
\bibitem [{\citenamefont {Mroue}\ \emph
  {et~al.}(2019{\natexlab{c}})\citenamefont {Mroue}, \citenamefont {Boyle},
  \citenamefont {Lovelace}, \citenamefont {Szilagyi}, \citenamefont {Pfeiffer},
  \citenamefont {Zenginoglu}, \citenamefont {Kidder}, \citenamefont {Taylor},
  \citenamefont {Hemberger},\ and\ \citenamefont
  {Scheel}}]{mroue_abdul_2019_3311869}%
  \BibitemOpen
  \bibfield  {author} {\bibinfo {author} {\bibfnamefont {A.}~\bibnamefont
  {Mroue}}, \bibinfo {author} {\bibfnamefont {M.}~\bibnamefont {Boyle}},
  \bibinfo {author} {\bibfnamefont {G.}~\bibnamefont {Lovelace}}, \bibinfo
  {author} {\bibfnamefont {B.}~\bibnamefont {Szilagyi}}, \bibinfo {author}
  {\bibfnamefont {H.}~\bibnamefont {Pfeiffer}}, \bibinfo {author}
  {\bibfnamefont {A.}~\bibnamefont {Zenginoglu}}, \bibinfo {author}
  {\bibfnamefont {L.}~\bibnamefont {Kidder}}, \bibinfo {author} {\bibfnamefont
  {N.}~\bibnamefont {Taylor}}, \bibinfo {author} {\bibfnamefont
  {D.}~\bibnamefont {Hemberger}}, \ and\ \bibinfo {author} {\bibfnamefont
  {M.}~\bibnamefont {Scheel}},\ }\href {\doibase 10.5281/zenodo.3311869}
  {\enquote {\bibinfo {title} {Binary black-hole simulation sxs:bbh:0047},}\ }
  (\bibinfo {year} {2019}{\natexlab{c}})\BibitemShut {NoStop}%
\bibitem [{\citenamefont
  {Collaboration}(2019{\natexlab{a}})}]{sxs_collaboration_2019_3315664}%
  \BibitemOpen
  \bibfield  {author} {\bibinfo {author} {\bibfnamefont {S.}~\bibnamefont
  {Collaboration}},\ }\href {\doibase 10.5281/zenodo.3315664} {\enquote
  {\bibinfo {title} {Binary black-hole simulation sxs:bbh:1392},}\ } (\bibinfo
  {year} {2019}{\natexlab{a}})\BibitemShut {NoStop}%
\bibitem [{\citenamefont
  {Collaboration}(2019{\natexlab{b}})}]{sxs_collaboration_2019_3315705}%
  \BibitemOpen
  \bibfield  {author} {\bibinfo {author} {\bibfnamefont {S.}~\bibnamefont
  {Collaboration}},\ }\href {\doibase 10.5281/zenodo.3315705} {\enquote
  {\bibinfo {title} {Binary black-hole simulation sxs:bbh:1410},}\ } (\bibinfo
  {year} {2019}{\natexlab{b}})\BibitemShut {NoStop}%
\bibitem [{\citenamefont {{LIGO Scientific
  Collaboration}}(2011)}]{ligonoisecurve}%
  \BibitemOpen
  \bibfield  {author} {\bibinfo {author} {\bibnamefont {{LIGO Scientific
  Collaboration}}},\ }\href {http://dcc.ligo.org/LIGO-T0900288/public}
  {\enquote {\bibinfo {title} {{Advanced ligo anticipated sensitivity
  curves}},}\ } (\bibinfo {year} {2011})\BibitemShut {NoStop}%
\bibitem [{\citenamefont {Ohme}\ \emph {et~al.}(2011)\citenamefont {Ohme},
  \citenamefont {Hannam},\ and\ \citenamefont {Husa}}]{Ohme:2011zm}%
  \BibitemOpen
  \bibfield  {author} {\bibinfo {author} {\bibfnamefont {F.}~\bibnamefont
  {Ohme}}, \bibinfo {author} {\bibfnamefont {M.}~\bibnamefont {Hannam}}, \ and\
  \bibinfo {author} {\bibfnamefont {S.}~\bibnamefont {Husa}},\ }\href {\doibase
  10.1103/PhysRevD.84.064029} {\bibfield  {journal} {\bibinfo  {journal} {Phys.
  Rev.}\ }\textbf {\bibinfo {volume} {D84}},\ \bibinfo {pages} {064029}
  (\bibinfo {year} {2011})},\ \Eprint {http://arxiv.org/abs/1107.0996}
  {arXiv:1107.0996 [gr-qc]} \BibitemShut {NoStop}%
\bibitem [{\citenamefont {Lindblom}\ \emph {et~al.}(2008)\citenamefont
  {Lindblom}, \citenamefont {Owen},\ and\ \citenamefont
  {Brown}}]{Lindblom:2008cm}%
  \BibitemOpen
  \bibfield  {author} {\bibinfo {author} {\bibfnamefont {L.}~\bibnamefont
  {Lindblom}}, \bibinfo {author} {\bibfnamefont {B.~J.}\ \bibnamefont {Owen}},
  \ and\ \bibinfo {author} {\bibfnamefont {D.~A.}\ \bibnamefont {Brown}},\
  }\href {\doibase 10.1103/PhysRevD.78.124020} {\bibfield  {journal} {\bibinfo
  {journal} {Phys. Rev.}\ }\textbf {\bibinfo {volume} {D78}},\ \bibinfo {pages}
  {124020} (\bibinfo {year} {2008})},\ \Eprint {http://arxiv.org/abs/0809.3844}
  {arXiv:0809.3844 [gr-qc]} \BibitemShut {NoStop}%
\bibitem [{\citenamefont {{Kumar}}\ \emph {et~al.}(2014)\citenamefont
  {{Kumar}}, \citenamefont {{MacDonald}}, \citenamefont {{Brown}},
  \citenamefont {{Pfeiffer}}, \citenamefont {{Cannon}}, \citenamefont
  {{Boyle}}, \citenamefont {{Kidder}}, \citenamefont {{Mrou{\'e}}},
  \citenamefont {{Scheel}}, \citenamefont {{Szil{\'a}gyi}},\ and\ \citenamefont
  {{Zengino{\v{g}}lu}}}]{2014PhRvD..89d2002K}%
  \BibitemOpen
  \bibfield  {author} {\bibinfo {author} {\bibfnamefont {P.}~\bibnamefont
  {{Kumar}}}, \bibinfo {author} {\bibfnamefont {I.}~\bibnamefont
  {{MacDonald}}}, \bibinfo {author} {\bibfnamefont {D.~A.}\ \bibnamefont
  {{Brown}}}, \bibinfo {author} {\bibfnamefont {H.~P.}\ \bibnamefont
  {{Pfeiffer}}}, \bibinfo {author} {\bibfnamefont {K.}~\bibnamefont
  {{Cannon}}}, \bibinfo {author} {\bibfnamefont {M.}~\bibnamefont {{Boyle}}},
  \bibinfo {author} {\bibfnamefont {L.~E.}\ \bibnamefont {{Kidder}}}, \bibinfo
  {author} {\bibfnamefont {A.~H.}\ \bibnamefont {{Mrou{\'e}}}}, \bibinfo
  {author} {\bibfnamefont {M.~A.}\ \bibnamefont {{Scheel}}}, \bibinfo {author}
  {\bibfnamefont {B.}~\bibnamefont {{Szil{\'a}gyi}}}, \ and\ \bibinfo {author}
  {\bibfnamefont {A.}~\bibnamefont {{Zengino{\v{g}}lu}}},\ }\href {\doibase
  10.1103/PhysRevD.89.042002} {\bibfield  {journal} {\bibinfo  {journal}
  {\prd}\ }\textbf {\bibinfo {volume} {89}},\ \bibinfo {eid} {042002} (\bibinfo
  {year} {2014})},\ \Eprint {http://arxiv.org/abs/1310.7949} {arXiv:1310.7949
  [gr-qc]} \BibitemShut {NoStop}%
\end{thebibliography}%
\end{document}